%% file: sample-base.tex
\documentclass[sigconf]{acmart}

\AtBeginDocument{%
  \providecommand\BibTeX{{%
    \normalfont B\kern-0.5em{\scshape i\kern-0.25em b}\kern-0.8em\TeX}}}

\usepackage{graphicx}
\usepackage{subcaption}
\usepackage{algorithm}
\usepackage[noend]{algpseudocode}
\usepackage{pifont}
\usepackage{multirow}
\usepackage{tabularx}
\usepackage[T1]{fontenc}
\begin{document}

\title{\textsc{Prompt4Vis}: Prompting Large Language Models with Example Mining and Schema Filtering for Tabular Data Visualization}
\author{Shuaimin Li}
\email{shuaimin.li@polyu.edu.hk}
\affiliation{%
  \institution{The Hong Kong Polytechnic University}
  \city{Hong Kong}
  \country{China}
}
\author{Xuanang Chen}
\email{chenxuanang19@mails.ucas.ac.cn}
\affiliation{%
  \institution{University of Chinese Academy of Sciences}
  \city{Beijing}
  \country{China}
}
\author{Yuanfeng Song}
\email{songyf@cse.ust.hk}
 \affiliation{
   \institution{The Hong Kong University of Science and Technology \& WeBank Co., Ltd}
   \city{Hong Kong}
   \country{China}}
   
\author{Yunze Song}
\email{YunzeSong77@gmail.com}
\affiliation{%
  \institution{University of Liverpool}
  \city{Liverpool}
  \country{UK}
}

\author{Chen Zhang}
\email{jason-c.zhang@polyu.edu.hk}
\affiliation{%
  \institution{The Hong Kong Polytechnic University}
  \city{Hong Kong}
  \country{China}
}
\input{abstract.tex}





\maketitle
\input{introduction}
\input{background}
\input{method}

\input{experiments}
\input{related_work}

\input{conclusion}
\bibliographystyle{ACM-Reference-Format}
\bibliography{sample-base}
\end{document}

%% file: abstract.tex
\begin{abstract}
   Data visualization (DV) systems are increasingly recognized for their profound capability to uncover insights from vast datasets, gaining attention across both industry and academia. Crafting data queries is an essential process within certain declarative visualization languages (DVLs, e.g., Vega-Lite, EChart.). The evolution of natural language processing (NLP) technologies has streamlined the use of natural language interfaces to visualize tabular data, offering a more accessible and intuitive user experience. However, current methods for converting natural language questions into data visualization queries, such as Seq2Vis, ncNet, and RGVisNet, despite utilizing complex neural network architectures, still fall short of expectations and have great room for improvement.
   
   Large language models (LLMs) such as ChatGPT and GPT-4, have established new benchmarks in a variety of NLP tasks, fundamentally altering the landscape of the field. Inspired by these advancements, we introduce a novel framework, \textsc{Prompt4Vis}, leveraging LLMs and in-context learning to enhance the performance of generating data visualization from natural language. \textsc{Prompt4Vis} comprises two key components: (1) a multi-objective example mining module, designed to find out the truly effective examples that strengthen the LLM's in-context learning capabilities for text-to-vis; (2) a schema filtering module, which is proposed to simplify the schema of the database. Extensive experiments through 5-fold cross-validation on the \textit{NVBench} dataset demonstrate the superiority of \textsc{Prompt4Vis}, which notably surpasses the state-of-the-art (SOTA) RGVisNet by approximately 35.9\% and 71.3\% on dev and test sets, respectively. To the best of our knowledge, \textsc{Prompt4Vis} is the first work that introduces in-context learning into the text-to-vis for generating data visualization queries.
\end{abstract}

%% file: introduction.tex
\section{Introduction}
In the modern era, big data serves as the primary driving force across various sectors. The analysis of big data to uncover underlying patterns is increasingly critical~\cite{LuoQ0018,QianRDKKMLC21,VartakHSMP16}. 
Data visualization emerges as a powerful tool in realizing this objective. 
Therefore, the topic of automatic data visualization has captured growing interest within the database and data mining communities~\cite{SavvidesHOP19,QianRDKKMLC21,VartakHSMP16}. For example, RGVisNet \cite{SongZWJ22} in KDD'22 proposes a hybrid retrieval-revision framework towards automatically data visualization generation.

One essential step in conducting data visualization is the formulation of visualization specifications using declarative visualization languages (DVLs), such as Vega-Lite~\cite{Vega}, ggplot2~\cite{ggplot2}, ZQL~\cite{SiddiquiKLKP16}, ECharts~\cite{LiMSSZWZC18}, and VizQL~\cite{Hanrahan06}. 
However, this composing specification demands users possess a thorough understanding of domain-specific data and familiarity with the syntax of these languages, which presents a significant challenge, especially for beginners. 
Thus, visualizing tabular data using natural language (or \textit{text-to-vis}), which aims to directly transform natural language questions (NLs) to data visualization language (DVs)~\cite{LuoQ0018,LuoQ00W18,Luo00CLQ21}, has garnered more and more attention within the community.

In an automatic text-to-vis system, it must first have a deep understanding of the natural language question and its corresponding database schema, then it needs to answer the given question with the correct data visualization language.
To achieve this goal, a series of efforts~\cite{CuiZWHCFZLZ20,DibiaD19,LuoQ0018,Luo00CLQ21,NarechaniaSS21} have been made, such as DeepEye~\cite{LuoQ0018}, NL4DV~\cite{NarechaniaSS21}, Seq2Vis~\cite{Luo00CLQ21}, ncNet~\cite{LuoTLTCQ22} and RGVisNet~\cite{SongZWJ22}.
Specifically, DeepEye and NL4DV rely on rule-based methodology or semantic parsing techniques in natural language processing, Seq2Vis~\cite{Luo00CLQ21} and ncNet~\cite{LuoTLTCQ22} attempt to build encoding-decoding frameworks using deep neural networks for text-to-vis, and RGVisNet~\cite{SongZWJ22} is retrieval-and-generation combined framework for data visualization language generation inspired by the concept of code reuse.
Although these efforts have achieved a noticeable enhancement in the performance of text-to-vis, such performance still falls short of expectations, especially when dealing with test data from different domains.

Recently, large language models (LLMs), especially the GPT series~\cite{BrownMRSKDNSSAA20,ChowdheryNDBMRBCSGSSTMRBTSPRDHPBAI23,openai2022,openai2023}, have revolutionized the field of natural language processing (NLP).
Leveraging their huge number of parameters and training data, LLMs learn substantial world knowledge and perform in pairs with humans ~\cite{abs-2302-04023,QinZ0CYY23}. Meanwhile, with the development of LLMs, in-context learning that does not rely on large-scale labeled data and does not require parameter updates~\cite{abs-2302-11042,abs-2301-13379,abs-2112-00114,Wei0SBIXCLZ22} attracts researchers in various fields. In-context learning enables LLMs to make predictions about a new example by learning from only a few labeled examples.
Hence, it is feasible and promising to leverage LLMs to realize text-to-vis and also to effectively alleviate the problem of insufficient generalization ability of existing methods.

To this end, this work attempts to adapt LLMs to generate data visualization queries from natural language questions and also proposes a novel prompting framework called \textbf{\textsc{Prompt4Vis}} with an in-context learning strategy.
In order to maximize the ability of LLMs on the text-to-vis, this work introduces two key components aimed at generating clearer and more effective prompt text for LLMs.
Firstly, as LLMs are sensitive to the choice of examples provided in the context~\cite{LuBM0S22,LiuSZDCC22}, an \textbf{example mining module} is designed to find out truly effective demonstrations, wherein the \textit{similarity} between the candidate examples and the target example, the \textit{influence} of candidate examples on the target example, and the \textit{diversity} among the candidate examples are all comprehensively considered, which indeed helps LLMs know what and how to perform well on this task. 
Specifically, the Euclidean distance between examples and the target input based on sentence vector representation is employed to measure the similarity and diversity of examples, and a contrastive learning-based influence model is trained to bring examples related to positive influence closer and push away from negative examples, this model can effectively measure the influence of examples.
In the process of selecting the prompt example set, we attempt to maximize the similarity and influence scores between the candidate example set and the target example while ensuring diversity within the candidate example set.
Secondly, given that encoding the entire schema for databases with numerous columns is not only expensive and impractical but also introduces irrelevant information to increase the difficulty of selecting the correct data for LLMs, a \textbf{schema filtering module} is also proposed to simplify the schema of database.
Specifically, considering the all-round capabilities of LLMs, we prompt to LLMs via in-context examples to help us select the necessary table in the schema of database for the input question, eliminating irrelevant and redundant schema related to the current question.

Extensive experiments are conducted on the widely-used multi-domain dataset \textit{NVBench}~\cite{Luo00CLQ21} to empirically verify the effectiveness of our \textsc{Prompt4Vis} framework.
Evaluation results demonstrate that \textsc{Prompt4Vis} not only outperforms all baseline methods with obvious improvements but also shows better stability across different cross-domains.
In summary, our contributions are as follows:
\begin{itemize}
    \item To the best of our knowledge, we are the first to introduce LLMs and in-context learning to generate data visualization for text-to-vis task. We believe this novel approach brings new insights to the text-to-vis task and will inspire further exploration of this promising new paradigm.
    \item We propose a novel framework \textsc{Prompt4Vis} for the adaption of LLMs on tabular data visualization task, with two novel example mining and schema filtering components.  
    \item  
    Extensive experiments demonstrate that \textsc{Prompt4Vis} bring around 36\% and 71\% relative improvements in overall accuracy on dev and test sets, respectively. Ablation studies also verify the effectiveness of all designed components in \textsc{Prompt4Vis}.
\end{itemize}
The rest of this paper is organized as follows: First, we introduced the background knowledge in section \ref{section:bg} to help realize this work. Then, section \ref{section:method} gives the details of the proposed \textsc{Prompt4Vis}. Next, we present the experimental results and discuss the findings in section \ref{section:exper}. Finally, the related work and conclusions of our work are introduced in section \ref{section:rw} and \ref{section:cl}.

%% file: background.tex
\section{Background}\label{section:bg}
\begin{figure}[t!]
  \centering
  \includegraphics[width=0.7\linewidth]{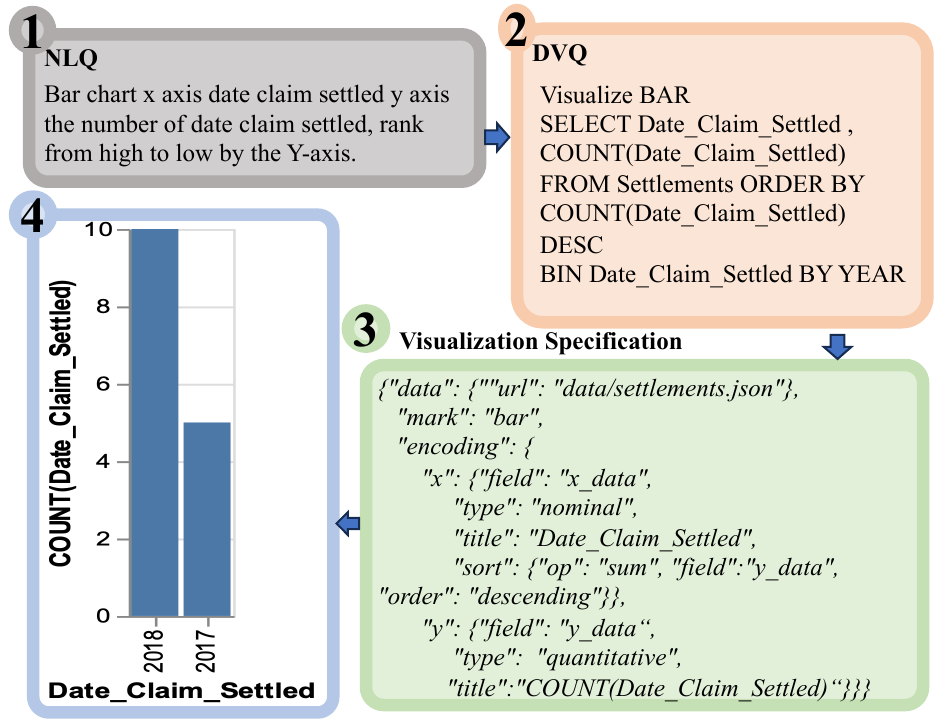}
  \caption{Pipeline of tabular data visualization.}
  \label{fig:background_example}
\end{figure}

In this section, we introduce the preliminary concepts including the Natural Language Question and the Visualization Specification, and then we give the task definition to help realize the work. 

\smallskip
\noindent \textbf{Natural Language Question (NLQ)} is a human-understandable expression used to describe the desired data visualization (DV), making it more user-friendly, particularly for novice users and those without a background in data science. 

\smallskip
\noindent \textbf{Visualization Specification} usually follows the grammar of a common declarative visualization language (DVL), pointing out the details of visualization, eg., data mapping, chart typologies, stylistic configurations, interactivity features, and layout design. Utilizing advanced DVLs, e.g., Vega-Lite \cite{Vega}, ggplot2 \cite{ggplot2}, ECharts \cite{LiMSSZWZC18}, the precise control of data visualization is achieved.

\smallskip
\noindent \textbf{Data Visualization Query (DVQ)} is proposed by Luo et al. \cite{LuoQ00W18,LuoQ0018}, involves initially executing a query on a database to retrieve the desired data, followed by defining the visualization details for presenting the acquired data. Importantly, DVQs are not limited to a single declarative visualization language (DVL). On the contrary, once the DVQ corresponding to a specific question is obtained, it can be seamlessly transformed to suit any DVL.

\smallskip
\noindent \textbf{Text-to-Vis} aims to translate NLQs into DVQs, which is a general step for tabular data visualization, as shown in Figure \ref{fig:background_example}. Formally, given an NLQ $q$ and corresponding database schema $s$, the text-to-vis task aims to character the corresponding DVQ $v$ to answer $q$. Specifically, schema $s$ consists of a collection of tables $\mathcal{T}_s = \{t_1,t_2,...,t_{n_T}\}$, where $t_i$ represents $i_{th}$ table in $\mathcal{T}_s$, $n_T$ is the number of tables in $T_s$. And $t_i$ is composed of a collection of columns, i.e., $t_i = \{c_1,c_2,...,c_{n_t}\}$, where $n_t$ is the number of columns in $t_i$.  

%% file: method.tex
\section{\textsc{Prompt4Vis}}\label{section:method}
In this section, we are ready to describe the proposed prompt framework to LLMs for tabular data visualization.

\begin{figure}[t!]
  \centering
  \includegraphics[width=\linewidth]{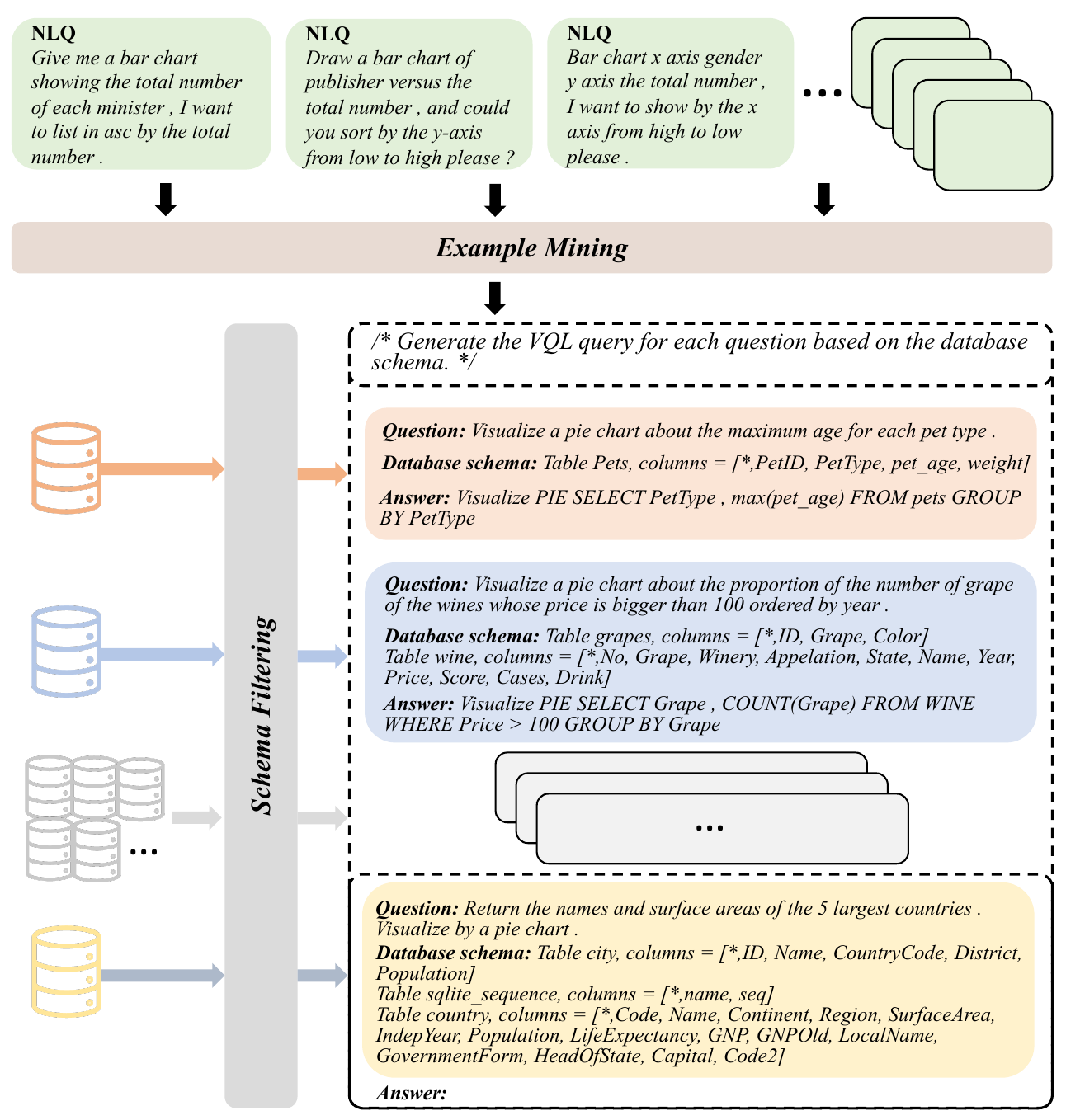}
  \caption{Workflow of \textsc{Prompt4Vis}, which prompts LLMs with an example mining module and a schema filtering module. The first module finds the truly effective examples for text-to-vis, and the second one simplifies the database schema.}
  \label{fig:prompt}
\end{figure}

\subsection{Overview}
In in-context learning, the model is initially presented with a set of labeled examples as input, then it predicts the output for new examples based on the labeled examples. For text-to-vis, each labeled example can be formalization as a triple: $(question, schema, dvq)$, where $question$ is the natural language question asked by users, $schema$ is the corresponding database for $question$, and $dvq$ denotes the target DVQ. Therefore, given a dataset $\mathcal{D} = {(q_n,s_n,v_n)}\sum_{n=1}^{N}$ containing $N$ $(question, schema, dvq)$ triples, a target question $q_t$, and a LLM $g$, in-context learning to generated DVQ $v_t$ based on database schema $s_t$ can be formulated as:

\begin{equation}
    v_t = g(\mathcal{P},(q_t,s_t))
\end{equation}\label{eq:icl}
where $\mathcal{P}$ is called prompt, which consists of $K$ examples, namely, $\mathcal{P}=\{(q_{1},s_{1},v_{1}),(q_{2},s_{2},v_{2}),...,(q_{K},s_{K},v_{K})\}$. 
Additionally, in-context learning also provides some context as instructions to prompt the model to achieve the ideal output.

Figure~\ref{fig:prompt} shows the workflow of \textsc{Prompt4Vis}.
As shown in Figure ~\ref{fig:prompt}, \textsc{Prompt4Vis} first finds effective examples with the example mining module, and then it conducts schema filtering to reduce the irrelevant information in the database. Finally, effective examples with filtered schemas are input into the LLMs to generate DVQs.

\subsection{Example Mining}
Previous research in in-context learning has demonstrated that similar examples of the target example can bring good performance \cite{zhang2023VisualPromptRetrieval,LiuSZDCC22}. However, they ignored the direct task-oriented influence of each example. To mitigate this gap, in addition to similarity, we propose to introduce influence in the process of example mining. At the same time, we introduce diversity to reduce the irrelevant information brought by similar candidate examples. In summary, 
our example mining method optimizes three different dimensions: influence, similarity, and diversity, to find out the most effective examples for text-to-vis.  

In the following, we will first introduce the methodology for calculating the metrics of influence, similarity, and diversity in section \ref{section:icv}. Then, we introduce the concentrate algorithm to find effective examples based on the aforementioned three measurements in section \ref{section:ESMA}. Since there are no existing methods to directly calculate the influence scores, we design an influence model and introduce the details about it in section \ref{section:influ_model}. 

In particular, in in-context learning for text-to-vis, each prompt example consists of a question, a database schema, and a data visualization query. Since the goal of this task is to answer a given natural language question, with the data schema serving merely as auxiliary information, and considering the significant differences among data schemas of different databases, we opt to select examples based on questions as the unit.

\subsubsection{\textbf{Methodology for Calculating Metrics}}\label{section:icv}
In this section, we define $\mathcal{C}(q_i)$ to measure the similarity of $q_i$ in $\mathcal{D}$ with the target question $q_t$, $\mathcal{I}(q_i)$ to measure the rewards influence of $q_i$. In the following, we introduce the detailed definitions and explanations of $\mathcal{C}(q_i)$, $\mathcal{I}(q_i)$, and $\mathcal{V}(A)$.
\paragraph{\noindent\textbf{Similarity $\mathcal{C}$.}}
The similarity metric $\mathcal{C}(q_i)$ is defined:
\begin{equation}
        \mathcal{C}(q_i) = \sum_{q_i\in\mathcal{D}}(1 - \frac{\rho(q_i,q_t) - \rho(q_j,q_t)}{\rho(q_l,q_t) - \rho(q_i,q_t)})
\end{equation}
where $\rho(q_i,q_t)$ measures the Euclidean distance of the vector representations between one single candidate question $q_i$ and the target question $q_t$. $\frac{\rho(q_i,q_t) - \rho(q_j,q_t)}{\rho(q_l,q_t) - \rho(q_i,q_t)}$ is the min-max normalization. $j$ is the index of the question that minimizes $\rho$:
\begin{equation}
    j = \underset{j \in |D|}{\mathrm{arg\,min}}\rho(q_j,q_t)
\end{equation}
and $l$ is the index of the question that maximizes $\rho$:
\begin{equation}
    l = \underset{l \in |D|}{\mathrm{arg\,max}}\rho(q_l,q_t)
\end{equation}

\paragraph{\noindent\textbf{Influence $\mathcal{I}$.}}
In relation to similarity measurement, we define influence sore $\mathcal{I}(q_i)$ as below:
\begin{equation}
        \mathcal{I}(q_i) = \sum_{q_i\in\mathcal{D}}\frac{\omega(q_i,q_t) - \omega(q_j,q_t)}{\omega(q_l,q_t) - \omega(q_i,q_t)}
\end{equation}
where $\omega(q_i,q_t)$ denotes the influence score of the candidate question $q_i$ to $q_t$. The evaluation of $\omega(.,.)$ is carried out with an influence model. We will go through the details of the designed influence model in section \ref{section:influ_model}.

\paragraph{\noindent\textbf{Diversity $\mathcal{V}$.}}
We define the diversity measurement $\mathcal{V}(A)$ as:
\begin{equation}
     \mathcal{V}(A) = \sum_{q\in\mathcal{D}}\rho(q_i,q_j)
\end{equation}
where $\rho(q_i,q_j)$ is the Euclidean distance between one single candidate question $q_i$ and another candidate question $q_t$. The greater the distance, the less similar the examples are to each other, indicating a higher diversity in the example subset.

\subsubsection{\textbf{Example Subset Mining Algorithm}}\label{section:ESMA}
The goal of example mining is to automatically select $K$ effective examples from the training dataset for a target question $q_t$. 

To achieve this goal, we employ an algorithm, Example Subset Mining (ESM) to find $K$ effective examples, it is composed of two phases including sorting and selecting. Initially, it considers only two criteria: similarity and influence, combing them for weight sorting. Subsequently, it further selects examples from the sorted list to maximize the diversity metric.

In other words, in the first step, the objective of example mining is to sort the training set $\mathcal{D}$; Second, we find an effective example subset $A$ from the sorted training set $\mathcal{D}_{sorted}$ by maximizing the diversity in it. 

Formally, the sorting process is implemented based on the score function $\phi$:
\begin{equation}
    \mathcal{D}_{sorted} = sort(\mathcal{D},\phi)
\end{equation}\label{eq:sort}
where $\phi$ is designed to measure the similarity and influence of $\mathcal{D}$, and it can be formalized in the following:
\begin{equation}
    \phi(q_i) = \alpha \mathcal{C}(q_i) + (1-\alpha) \mathcal{I}(q_i)
\end{equation}
where $\mathcal{C}(q_i)$ measures the similarity of $q_i$ in $\mathcal{D}$ with the target question $q_t$, $\mathcal{I}(q_i)$ rewards influence of $q_i$.

Then, in the second step, an example subset $S$ is selected by maximizing the diversity objective:
\begin{equation}
        A^* = \underset{A:A \subseteq \mathcal{D}_{sorted}}{\mathrm{arg\,max}}\ \mathcal{V}(A) \text{ subject to } |A| \leq K
\end{equation}
where $\mathcal{V}(A)$ measures the diversity of $A$.

To this end, Example Subset Mining (ESM) Algorithm is illustrated in Algorithm \ref{alg:GES}. 
ESM first searches over questions in the training set for target questions and then sorts them with function $\phi$. Next, the diversity of the examples is maximized to find Example Subset $A$.

\begin{algorithm}
\footnotesize
\caption{Example Subset Mining (ESM)}
\label{alg:GES}
\begin{algorithmic}[1] 
\Statex \textbf{Input}: Training Set $\mathcal{D}$, Example Number of Training Set $N$, Example Number of ExampleSubset $M$, Parameter $\alpha$, $\beta$.
\Statex \textbf{Onput}: Example Subset $A$
\Procedure{ExampleSubsetMining}{$parameters$} 
    \State Initialize an empty list $L$
    \For{each element $e$ in $\mathcal{D}$}
        \State Compute $score$ for $e$ with $\phi$;
        \State Append $(e,score)$ to $L$;
    \EndFor
    \State Sort $L$ by $score$ in descending order
    \State \( \mathcal{D}_{sorted} \gets \emptyset \)
    \For{each element $e$ in $L$}
        \State \( \mathcal{D}_{sorted} \gets \mathcal{D}_{sorted} \cup \{e\} \);
    \EndFor
    \State \( A \gets \emptyset \)
    \While{\( |A| < K \)}
        \State \( z' \gets \arg\max_{z\in\mathcal{D}_{sorted}-A}(\mathcal{V}(\{z\} \cup A) - \mathcal{V}(A)) \);
        \State \( A \gets A \cup \{z'\} \);
        \If{\( \mathcal{D}_{sorted} - A = \emptyset \)}
            \State \textbf{break};
        \EndIf
    \EndWhile
\State \textbf{return} Example Subset \( A \).
\EndProcedure
\end{algorithmic}
\end{algorithm}

\subsubsection{\textbf{Contrastive Learning-based Influence Model}}\label{section:influ_model}
\begin{figure*}[h]
  \centering
  \includegraphics[width=\linewidth]{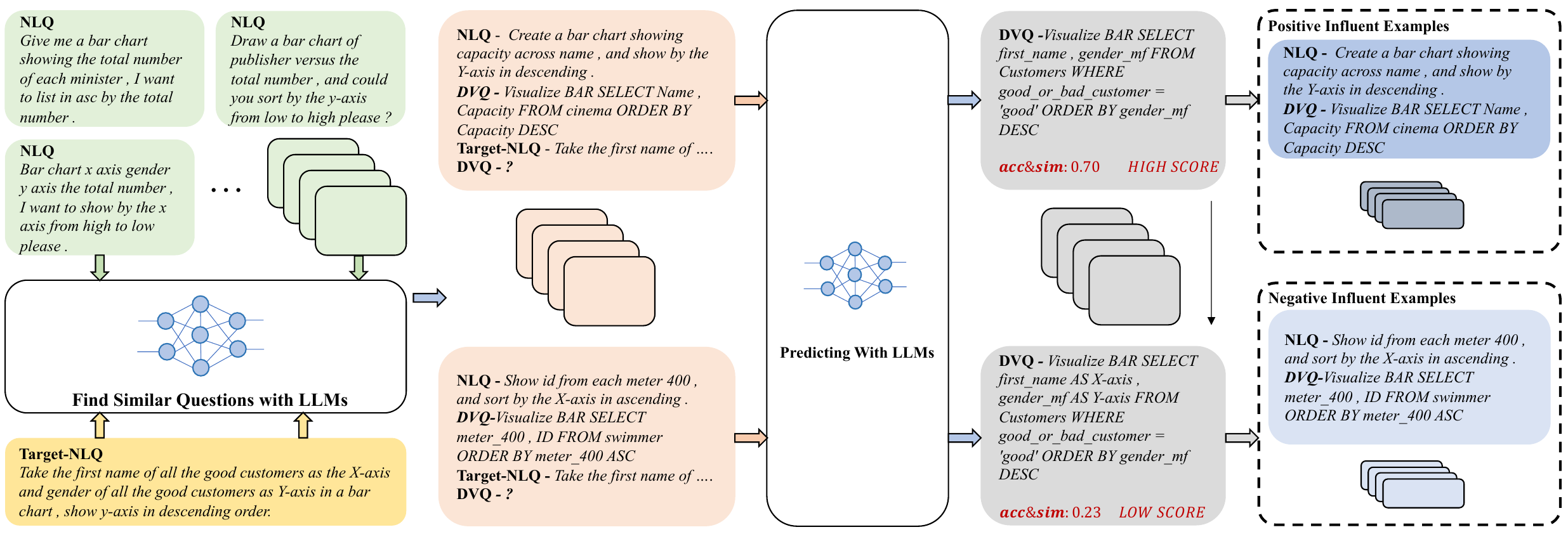}
  \caption{Training data construction for the influence model, which first finds similar questions with LLMs and takes each one of them as a prompt example. Then, these questions are sorted based on the influence scores of the DVQs generated by the LLMs. Finally, the positive and negative sets are established based on influence scores.}
  \label{fig:data_construction}
\end{figure*}
As mentioned above, similarity and diversity are realized through sentence vector representation and Euclidean distance calculation. However, the influence of each example can not obtained by the existing methods. Therefore we designed a contrastive learning-based influence learning model. 

In this section, we will provide a detailed introduction to the training data construction for the influence model and training objective.

\textbf{Training Data Construction for Influence Model.}
To train the contrastive learning-based influence model, we need to construct the training data in the first step. In other words, we need to find a positive and negative set for each example $q_i \in \mathcal{D}_{Train}^I$ in the training set $\mathcal{D}_{Train}^I$ of the influence model. Recall that in-context learning is defined as $v_t = g(\mathcal{P},(q_t,s_t))$ in Equation \ref{eq:icl}. For each example, given the question $q_n$ and the schema $s_n$, we predict $\hat{v_n}=g((q_m,s_m,v_m),(q_t,s_t))$ where $g$ is the large language model and $q_m$ is the question in the prompt example, $s_m$ is the corresponding database schema for $q_m$, noted that $m!=t$. In particular, $q_m$ is extracted from a set that includes the top-$L$ similar questions to $q_t$. Since we have the ground truth for $v_t$ for $(q_t,s_t)$, we can measure the performance by comparing the predicted $\hat{v_t}$ and ground truth $v_t$. Then the performance will be set as the influence score of $q_m$ on $q_t$ in in-context learning of text-to-vis. 

Concretely, the influence score consists of two aspects, i.e., the average accuracy of the predicted $\hat{v_t}$, and the semantic similarity between the predicted $\hat{v_t}$ and the ground truth $v_t$. Formally, it can be formalized as:
\begin{align}
    \omega(q_i,q_t) &= \lambda \textit{Sim}(\hat{v_t},v_t) + (1-\lambda)\textit{Acc}(\hat{v_t},v_t)\\
    \hat{v_t} &= g((q_i,s_i,v_i),(q_t,s_t))
\end{align}
where $\textit{Sim}(\hat{v_n},v_n)$ is measured by Euclidean distance function $-\rho$. $\textit{Acc}(\hat{v_n},v_n)$ is the average score of the four metrics described in section \ref{section:evaluation_metric}.

For each $q_n$, we extract the top-$M$ questions with the highest influence score and their corresponding database schema to form the positive examples, while the lowest $M$ as the negative examples. The workflow of training data construction for the influence model is shown in Figure \ref{fig:data_construction}. 

\textbf{Training Objective.}
In the influence model, we first encode the input questions with a pre-trained language model BERT \cite{DevlinCLT19}: $h = f_{\theta}(q)$, and then fine-tune the parameters using the contrastive learning objective. Inspired by \cite{GaoYC21}, we train the influence model to follow the contrastive framework with a cross-entropy objective. Let $h(q_i)$ denote the representation of target question, $q_i^+$ and $q_i^-$ denote the representations of positive and negative questions for $q_i$, then the training objective for $(q_i,q_i^+,q_i^-)$ within a mini-batch of $M$ pairs is:
\begin{equation}\label{eq:tau}
    -log\frac{e^{cos}(f(q_i),f(q_i^+))/\tau}{\sum_{j=1}^{N}(e^{cos(f(q_i),f(q_j^+))/\tau})+e^{cos(f(q_i),f(q_j^-))/\tau})}
\end{equation}
where $\tau$ is a temperature hyper-parameter and $cos(f(q_1),f(q_2))$ denotes the cosine similarity $\frac{f(q_1)^{\top} f(q_2)}{||f(q_1)||\cdot||f(q_2)||}$.

\subsection{Schema Filtering}
Apart from the natural language question $q_t$, the input in the prompt examples includes a database schema $s_t$ consisting of a set of tables $\mathcal{T}$, with each table $tb \in \mathcal{T}$ comprising of a set of columns $c$. 
The gold DVQ query $v_t$ for the target question mentions a subset $R(v_t)$ of schema elements from $s_t$. 
Schema elements can be either tables or columns. 
In the schema filtering process, we propose to filter a subset $\hat{R}(v_t)$ from $s_t$ covering $R(v_t)$, i.e., $R(v_t) \subseteq \hat{R}(v_t)$ and $|\hat{R}(vt)| << |s_t|$. In this work, we adopt a table as the basic unit in the schema filtering for two reasons. First, extracting precise column information is challenging because sometimes there is no direct semantic connection between the natural language description and the database column names. Additionally, at times, the operations of DVQ are performed across multiple column names in multiple tables. Providing relatively coarse-grained base units can help prevent the omission of information.

To be specific, we use state-of-the-art LLMs with few-shot prompting to produce $\hat{R}(v_t)$. We employ GPT-3.5-Turbo with a fixed prompt comprising of ten in-context examples as shown in Table \ref{tab:schema_filter_case} to create the desired $\hat{R}(v_t)$ corresponding to $q_t$.

\begin{table}
    \centering
    \footnotesize
    \begin{tabular}{|p{8cm}|}
        \toprule
         \textbf{Select the related tables for generating SQL queries for each question based on the database schema.}\\
         \midrule
         \textcolor{blue}{\textbf{Question:}} Give me a bar chart showing the total number of each minister , I want to list in asc by the total number . \\
         \textcolor{blue}{\textbf{Schema:}} Table region, columns = [*,Region\_ID, Region\_name, Date, Label, Format, Catalogue]\\
         Table party, columns = [*,Party\_ID, Minister, Took\_office, Left\_office, Region\_ID, Party\_name]\\
         Table member, columns = [*,Member\_ID, Member\_Name, Party\_ID, In\_office]\\
         Table party\_events, columns = [*,Event\_ID, Event\_Name, Party\_ID, Member\_in\_charge\_ID]\\
        \textcolor{blue}{\textbf{Selected Table:}} Table party\\
        \midrule
        ... \\
        \midrule
        \textcolor{blue}{\textbf{Question:}} Which catalog contents has price above 700 dollars ? Show their catalog entry names and capacities , list by the X in ascending . \\
        \textcolor{blue}{\textbf{Schema:}} Table Attribute\_Definitions, columns = [*,attribute\_id, attribute\_name, attribute\_data\_type]\\
        Table Catalog\_Structure, columns = [*,catalog\_level\_number, catalog\_id, catalog\_level\_name]\\
        Table Catalog\_Contents, columns = [*,catalog\_entry\_id, catalog\_level\_number, parent\_entry\_id, previous\_entry\_id, next\_entry\_id, catalog\_entry\_name, product\_stock\_number, price\_in\_dollars, price\_in\_euros, price\_in\_pounds, capacity, length, height, width]\\
        Table Catalog\_Contents\_Additional\_Attributes, columns = [*,catalog\_entry\_id, catalog\_level\_number, attribute\_id, attribute\_value]\\
        \textcolor{blue}{\textbf{Selected Table:}}\\
        \bottomrule
    \end{tabular}
    \caption{Prompt example in schema filtering module.}
    \label{tab:schema_filter_case}
\end{table}

%% file: experiments.tex
\section{Experiments}\label{section:exper}
\subsection{Experimental Setup}\label{section:ex_setup}

\subsubsection{Dataset}
In this work, we utilize \textit{NVBench}, a public text-to-vis dataset, to conduct our experiments. It is initially created to evaluate text-to-vis systems. In particular, \textit{NVBench} contains 7247 DVQs, each corresponding to several natural language questions and a specific database schema. Following the previous work \cite{SongZWJ22}, we partitioned the dataset with the training set, development set, and test set containing 98, 29, and 14 databases, respectively. 
Furthermore, to thoroughly validate the effectiveness of the experiments and mitigate the impact of different database partitions, we conducted a 5-fold cross-validation.

\subsubsection{Baselines}
In this work, we compare four widely recognized baselines with our method for evaluation, including Seq2Vis~\cite{Luo00CLQ21}, Transformer~\cite{VaswaniSPUJGKP17}, ncNet~\cite{LuoTLTCQ22}, and RGVisNet\cite{SongZWJ22}. 

\subsubsection{Evaluation Metrics}\label{section:evaluation_metric}
Following the prior works~\cite{SongZWJ22}, we use four popular metrics to evaluate the models in the experiments including \textit{Vis Accuracy}, \textit{Axis Accuracy}, \textit{Data Accuracy}, and \textit{Overall Accuracy}. 

\begin{table*}[t]
    \centering
    \begin{tabular}{l|rrrr|rrrr}
        \toprule
        &\multicolumn{4}{c}{Test Set}&\multicolumn{4}{c}{Dev Set}\\
        \midrule
        Method  & Vis Acc& Axis Acc & Data Acc & Overall Acc & Vis Acc& Axis Acc & Data Acc & Overall Acc\\
        \midrule
        Seq2Vis  &86.97\%&0.02\%&11.88\%&0.01\%&84.10\%&0.81\%&11.31\%&0.32\%\\
        Transformer &98.82\%&0.58\%&12.16\%&0.42\%&\textbf{98.31}\%&2.54\%&11.11\%&1.71\%\\
        ncNet  &\textbf{98.86\%}&41.34\%&40.62\%&23.61\%&98.27\%&37.44\%&45.79\%&23.97\%\\
        RGVisNet &95.46\%&44.80\%&37.35\%&30.75\%&95.38\%&60.06\%&52.37\%&44.44\%\\
        \midrule
        \textsc{Prompt4Vis} (ours)&98.37\%&\textbf{79.23\%}&\textbf{58.64\%}&\textbf{52.69\%}&97.77\%&\textbf{79.23\%}&\textbf{65.68\%}&\textbf{60.39\%}\\
        \bottomrule
    \end{tabular}
    \caption{Average results of the baseline model and our \textsc{Prompt4Vis} on the NVBench dataset using 5-fold cross-validation. }
    \label{tab:main_results}
\end{table*}

\subsection{Experimental Results}\label{section:ex_results}
\subsubsection{Main results} We first compare the proposed method with baselines on the aforementioned metrics in a 5-fold cross-validation set, and we report the average scores in Table \ref{tab:main_results}. As is shown, \textbf{\textsc{Prompt4Vis} outperforms the SOTA RGVisNet with obvious improvements.} Specifically, \textsc{Prompt4Vis} method surpasses the current SOTA RGVisNet by relatively 71.3\% and 35.9\% on test and dev set in terms of overall accuracy, respectively, which indicates the absolute effectiveness of the proposed method. This finding shows a novel paradigm of text-to-vis different from the previous designing of complex neural networks.
 
\begin{figure}[t]
    \centering
    \begin{subfigure}[b]{0.22\textwidth}
        \includegraphics[width=\textwidth]{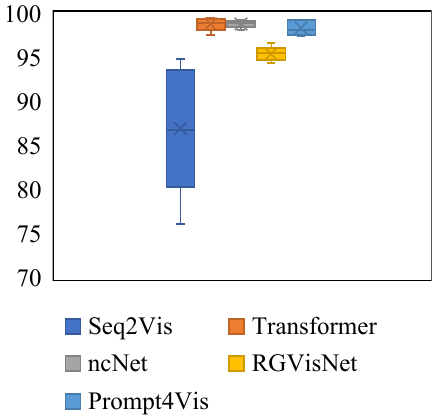}
        \caption{Vis Acc}
        \label{fig:sub1}
    \end{subfigure}
    \hfill 
    \begin{subfigure}[b]{0.22\textwidth}
        \includegraphics[width=\textwidth]{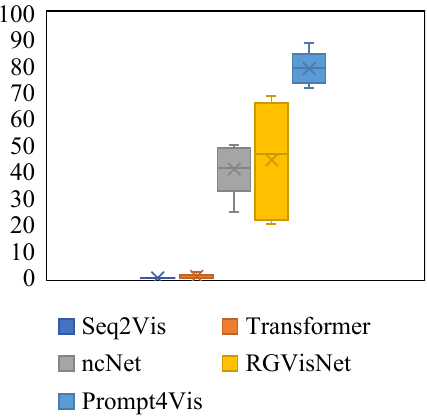}
        \caption{Axis Acc}
        \label{fig:sub2}
    \end{subfigure}
    \hfill
    \begin{subfigure}[b]{0.22\textwidth}
        \includegraphics[width=\textwidth]{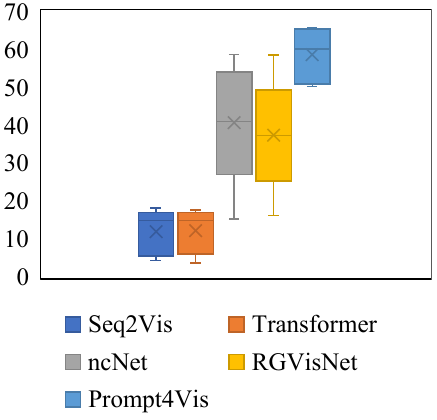}
        \caption{Data Acc}
        \label{fig:sub3}
    \end{subfigure}
    \hfill
    \begin{subfigure}[b]{0.22\textwidth}
        \includegraphics[width=\textwidth]{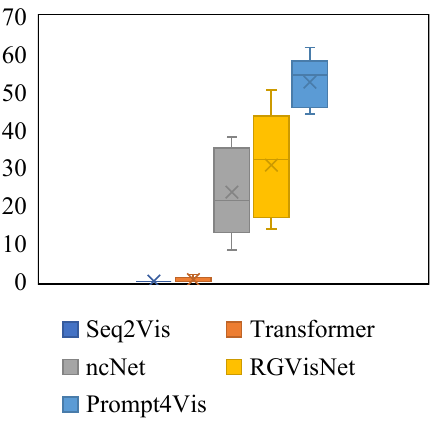}
        \caption{Overall Acc}
        \label{fig:sub4}
    \end{subfigure}
    \caption{Performance range in 5-fold cross-validation across various models and metrics.}
    \label{fig:robustness}
\end{figure}

Then, we present the performance range in 5-fold cross-validation across different models and metrics with box charts in Figure \ref{fig:robustness}. As is shown, \textbf{\textsc{Prompt4Vis} demonstrates better stability across domains.}
To be specific, when compared to other baseline models such as ncNet and RGVisNet, the box height for \textsc{Prompt4Vis} is consistently the shortest across various metrics. In other words, \textsc{Prompt4Vis} exhibits the least fluctuation in terms of all accuracy metrics. It indicates that our model has the best stability in cross-domain settings. Moreover, for each subplot, we can see that our model is always located in the top-right corner of the subplots, which means that the performance of \textsc{Prompt4Vis} consistently exceeds the baseline models.

However, we observed that although our model surpasses the current SOTA baseline model RGVisNet by relatively 35.9\% to 71.3\%, there remains considerable potential for enhancing performance in the text-to-vis task. This finding highlights the difficulty of text-to-vis and suggests further efforts are required in this field.

\begin{table*}
    \centering
    \begin{tabular}{l|rrrr|rrrr}
        \toprule
        &\multicolumn{4}{c}{Test Set}&\multicolumn{4}{c}{Dev Set}\\
        \midrule
        Method  & Vis Acc& Axis Acc & Data Acc & Overall Acc& Vis Acc& Axis Acc & Data Acc & Overall Acc \\
        \midrule
        \textsc{Prompt4Vis} (ours) &98.37\%&79.23\%&58.64\%&52.69\%&97.77\%&79.23\%&65.68\%&60.39\%\\
        \midrule
        w. random &90.68\%&58.67\%&45.09\%&33.51\%&89.97\%&59.67\%&53.72\%&39.65\% \\
        w. sim  &96.96\%&76.57\%&57.33\%&50.15\%&96.48\%&76.58\%&63.35\%&56.75\%\\
        \midrule
        w/o inf &97.72\%&77.01\%&57.64\%&50.72\%&97.12\%&77.73\%&64.33\%&58.75\%\\
        w/o div &97.22\%&78.61\%&58.25\%&51.65\%&97.11\%&78.36\%&64.18\%&58.32\%\\
        w/o sim &98.29\%&79.79\%&58.59\%&52.16\%&97.74\%&80.65\%&64.98\%&59.40\%\\
        \midrule
        w. all schemas&98.20\%&76.10\%&57.02\%&50.42\%&97.60\%&74.53\%&63.81\%&56.40\%\\
        w/o schema&98.68\%&76.69\%&57.76\%&50.78\%&97.99\%&76.97\%&63.45\%&57.34\%\\
        w. RAT&97.92\%&76.77\%&57.11\%&50.46\%&97.61\%&74.99\%&63.72\%&56.72\%\\
        w. RATLink&98.14\%&58.14\%&48.86\%&38.33\%&97.23\%&60.90\%&57.65\%&46.83\%\\
        \bottomrule
    \end{tabular}
    \caption{Ablation study results.}
    \label{tab:ablation_study}
\end{table*}

\subsubsection{Ablation Study}
We conduct experiments to verify the effectiveness of each component in \textsc{Prompt4Vis}. To be specific, we set  \textsc{Prompt4Vis} with all designed components as the baseline. Then we create variants of \textsc{Prompt4Vis} by either removing or replacing the designed components and then compare their performance with \textsc{Prompt4Vis}. The experimental results are presented in Table \ref{tab:ablation_study}. In the following, we will introduce the details of these variants and analyze their performance.

First, we investigate if the improvements of our method are brought by only the ability of large language models. Therefore, we propose to prompt GPT-3.5-Turbo with 5 random examples sampling from the training set and name it as \textbf{w. random}. Then, as prior studies demonstrate selecting similar examples can benefit in-context learning \cite{zhang2023VisualPromptRetrieval,LiuSZDCC22}. We propose a variant, \textbf{w. sim} as a baseline to verify the proposed multi-objective example mining method. As shown in Table \ref{tab:ablation_study}, \textbf{besides the superior ability of LLMs, our prompt method makes obvious contributions to the improvements.} In particular, \textsc{Prompt4Vis} outperforms LLMs prompting with random examples with large margins, 57.2\% and 52.3\% on test and dev set, respectively, which shows the effectiveness of our method. In other words, directly employing LLMs with random examples can not obtain the ideal performance. What's more, compared with the results in Table \ref{tab:main_results}, LLMs prompting with random examples even performs worse than the RGVisNet on the dev set. Additionally, the proposed method also achieves better performance than prompting with similar examples with obvious margins.

Second, we investigate the utility of three elements in an example mining module. We remove similarity, influence, and diversity elements, respectively, and name them \textbf{w/o sim}, \textbf{w/o influence}, and \textbf{w/o diversity}. As shown in Table \ref{tab:ablation_study}, experimental results show the effectiveness of each element in the example mining module, when removing any element in it, the performance drops.

Finally, we verify the effectiveness of the second module, schema filtering. We first remove the scheme filtering module, which means we provide full schemes for each example (\textbf{w. all schemas}). Then, we only provide the schema for the target question and name this method as \textbf{w/o schema}. Next, we replace the schema filtering with a representative schema linking method RAT \cite{WangSLPR20}, and give tables filtered by this method to LLMs (\textbf{w. RAT}). Finally, we also explore providing the schema linking, which means the concentrate columns that the target query may mention with RAT, and name it \textbf{w. RATLink}. As shown in Table \ref{tab:ablation_study}, we can see that \textbf{w. all schemas} hurts the performance, which drops relatively by 4.3\% on the test set and 6.6\% on the dev set compared with \textsc{Prompt4Vis}. The reason may be that a large number of tables bring irrelevant information to the model. What's more, when replacing the schema filtering module proposed in this work with the RAT method, the performances also drop around 4.2\% to 6.1\%, which further indicates the effectiveness of the schema filtering module in \textsc{Prompt4Vis}.

\subsubsection{Parameter Study.} 
\begin{figure}[t]
    \centering
    \begin{subfigure}[b]{0.22\textwidth}
        \includegraphics[width=0.8\textwidth]{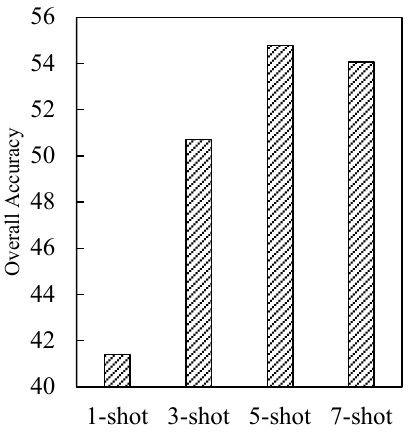}
        \caption{Test set.}
        \label{fig:sub1_number}
    \end{subfigure}
    \begin{subfigure}[b]{0.22\textwidth}
    \includegraphics[width=0.8\textwidth]{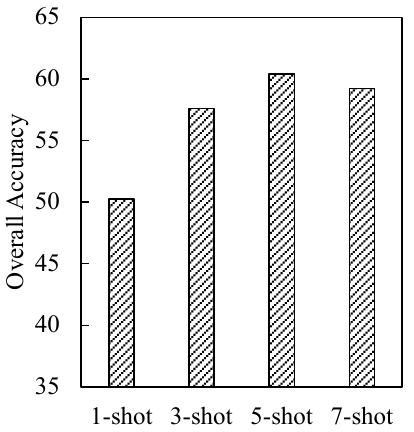}
    \caption{Dev set.}
    \label{fig:sub2_number}
    \end{subfigure}
    \caption{Overall accuracy with the different number of in-context examples.}
    \label{fig:number_in_context}
\end{figure}

In this section, we investigate the number of in-context examples for \textsc{Prompt4Vis}. Following the approach of \cite{BarGDGE22}, we vary the number of in-context examples from 1 to 7 and test the performance of \textsc{Prompt4Vis} on the test sets of 5-fold cross-validation setting. The average scores of overall accuracy are presented in Figure \ref{fig:number_in_context}. The experimental results demonstrate that more examples can provide richer context, thereby enhancing model performance. However, when the number of examples reaches a certain threshold, such as 7 examples, the model performance is similar and even lower than 5 examples. This may be because too many examples could bring irrelevant information, implying that in in-context learning, choosing an optimal number of sample examples can result in a higher cost-effectiveness ratio rather than indiscriminately increasing the number of examples.

\subsubsection{Case Study}
\begin{table*}
    \centering
    \begin{tabular}{l|p{6cm}|l}
        \toprule
        NL Question  & Return the names and surface areas of the 5 largest countries. Visualize by a pie chart. & \multirow{3}{*}{
        \begin{subfigure}[b]{0.23\textwidth}
        \caption{Seq2Vis and Transformer \ding{56}}
        \includegraphics[width=0.5\textwidth]{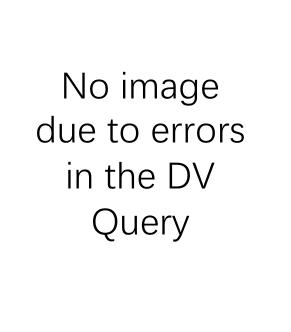}
        \label{fig:sub1_case_pic}
        \end{subfigure}
        \begin{subfigure}[b]{0.23\textwidth}
        \caption{ncNet \ding{56}}
        \includegraphics[width=0.9\textwidth]{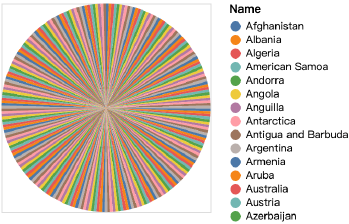}
        \label{fig:sub2_case_pic}
        \end{subfigure}}\\ \cline{1-2}
        Target DVQ & VISUALIZE PIE SELECT Name, SurfaceArea FROM country ORDER BY SurfaceArea DESC LIMIT 5 &\\\cline{1-2}
        Seq2Vis &VISUALIZE BAR SELECT Surface, COUNT(*) FROM hiring GROUP BY Surface ORDER BY surface ASC $\rightarrow$ \textit{Figure (a)} &\\ 
        Transformer & VISUALIZE PIE SELECT Area, COUNT(Area) FROM appellations GROUP BY Area $\rightarrow$ \textit{Figure (a)}& \multirow{4}{*}{
        \begin{subfigure}[b]{0.23\textwidth}
        \caption{RGVisNet \ding{56}}
        \includegraphics[width=0.95\textwidth]{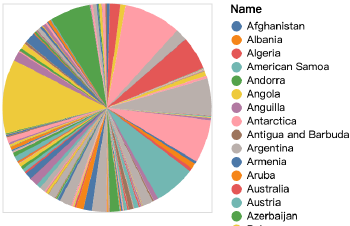}
        \label{fig:sub3_case_pic}
        \end{subfigure}
        \begin{subfigure}[b]{0.23\textwidth}
        \caption{\textsc{Prompt4Vis} \ding{52}}
        \includegraphics[width=0.91\textwidth]{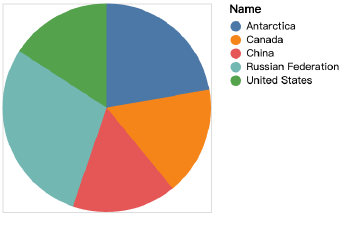}
        \label{fig:sub4_case_pic}
        \end{subfigure}}\\
        ncNet &VISUALIZE PIE SELECT Name, COUNT(*) FROM country $\rightarrow$ \textit{Figure (b)}&\\
        RGVisNet &VISUALIZE PIE SELECT Name, SurfaceArea FROM country $\rightarrow$ \textit{Figure (c)}&\\
       \textsc{Prompt4Vis} (ours)&VISUALIZE PIE SELECT Name, SurfaceArea FROM country ORDER BY SurfaceArea DESC LIMIT 5 $\rightarrow$ \textit{Figure (d)}&\\
        \bottomrule
    \end{tabular}
    \caption{Case Study. DVQs generated by baselines and \textsc{Prompt4Vis} and their corresponding charts.}
    \label{tab:case_study}
\end{table*}

Table \ref{tab:case_study} presents a case to concretely show the DVQs generated by \textsc{Prompt4Vis} and the baselines. The corresponding charts generated by these models are also displayed in Table \ref{tab:case_study}. As shown in Table \ref{tab:case_study}, Seq2Vis and Transformer produce wrong table names, which leads to no image generated in Table \ref{fig:sub1_case_pic}. ncNet produce the DVQ with correct table \textit{country}, however, it selects \textit{COUNT(*)} from \textit{country}, which results in a wrong image in Table \ref{fig:sub2_case_pic}. The first half of the DVQ generated by RGVisNet is completely correct, as it can learn the prototype close to the gold query by retrieval and obtain the DVQ for the current NLQ through revision. However, it lacks \textit{ORDER BY SurfaceArea DESC LIMIT 5}, which may be due to the rarity of this expression in the training set, not enabling the neural network to learn this pattern. Different from the aforementioned models, \textsc{Prompt4Vis} is able to accurately produce the DVQ as the same as the target query, which results in the correct charts presented in Table \ref{fig:sub4_case_pic}.

%% file: related_work.tex
\section{Related Work}\label{section:rw}
\subsection{In-context Learning}
Recently, LLMs, represented by ChatGPT and GPT-4, have demonstrated remarkable ability in comprehending natural language and have made significant advancements in a series of NLP tasks \cite{abs-2302-04023,QinZ0CYY23}.
In-context learning is a novel learning-free paradigm that originally emerged in LLMs. Recently, it has inspired the passion of database researchers \cite{NarayanCOR22,abs-2307-07306,abs-2304-11015,abs-2308-15363}. For instance, Narayan et al. \cite{NarayanCOR22} explore the ability of in-context learning in data cleaning and integration tasks including error detection, entity matching, and data transformation. Dong et al. \cite{abs-2307-07306} propose a zero-shot method for text-to-SQL, named C3, which first formulates clear task prompts, then devises calibration techniques to address incorrect outputs, and finally orders the large language model to produce consistent outputs to enhance the quality of the generated SQL. 

Different from the above works, our work mainly focuses on text-to-vis and explores how to find effective examples for in-context learning of text-to-vis.

What's more, researchers in NLP and CV have found that the examples selected of in-context learning have a great influence on the performance of downstream tasks~\cite{LiuSZDCC22,zhang2023VisualPromptRetrieval}. They have made efforts to find good examples of in-context learning. For instance, Liu et al.~\cite{LiuSZDCC22} supposed that good examples should be semantically similar to the target example. Rubin et al.~\cite{RubinHB22} and Zhang et al.~\cite{zhang2023VisualPromptRetrieval} retrieve examples that can maximize the performance of downstream tasks for in-context learning. Different from the aforementioned works, we find effective examples from three different dimensions for text-to-vis and conduct schema filtering to give clean prompts to LLMs.

\subsection{Text-to-Vis}
Text-to-vis has garnered significant interest from the database and visualization communities as it enables non-experts to interact with the visualization systems through natural language questions. Existing typical works for text-to-vis can be divided into rule-based methods and neural network-based methods.

Rule-based methods represent the main trend at the beginning of text-to-vis. For instance, a rule-based approach to transform textual commands into infographics is employed in text-to-vis by Cui et al. \cite{CuiZWHCFZLZ20}. Moritz et al. \cite{MoritzWNLSHH19} design a set of constraints to model the knowledge in visualization and then optimize these constraints. DeepEye \cite{LuoQ0018} is an automatic visualization system that employs semantic parsing tools in NL. It consists of three components including visualization recognition, visualization ranking, and visualization selection. NL4DV \cite{NarechaniaSS21} is also implemented based on parsing tools and offers a Python toolkit that supports various high-level operations to assist users in creating NL-based DV systems. 

To further promote the field of text-to-vis, Luo et al.~\cite{Luo00CLQ21} create a cross-domain text-to-vis dataset, \textit{NVBench}, based on a popular text-to-SQL benchmark. Meanwhile, with the development of neural networks, the Seq2Vis model~\cite{Luo00CLQ21} is proposed based on the encoder-decoder framework and neural networks. 
Inspired by the way developers reuse previously validated code snippets from code search engines or large codebases during software development, Song et al.~\cite{SongZWJ22} introduce a novel hybrid retrieval-generation framework for text-to-vis called RGVisNet. It retrieves the most relevant DVQ candidate as a prototype from the DVQ codebase and then refines this prototype to generate the desired DVQ.

Most recently, visualization researchers have also been attracted by large language models, which further promote the development of data visualization \cite{MaddiganS23,Dibia23,WangTL24,abs-2309-10245,WangZWLW23}. For instance, Chat2Vis system~\cite{MaddiganS23} makes efforts to prompt LLMs to produce Python code for data visualization. LIDA~\cite{Dibia23} is a tool to visualize with the assistance of large language models and image generation models, which provides a novel Python API and a user interface to interact with users. Ko et al. \cite{abs-2309-10245} aim at promoting the development of visualization by enriching the datasets in this field. Therefore, they propose a novel framework based on LLMs to generate natural language datasets taking Vega-Lite specifications. What's more, Wang et al.~\cite{WangZWLW23} propose to leverage LLMs as a recommendation tool for visualization.

Several of the aforementioned methods utilize LLMs as integral components within their frameworks for various visualization tasks, including recommendation~\cite{WangZWLW23} and concept binding~\cite{WangTL24}. Additionally, some approaches leverage LLMs for generating programming code~\cite{MaddiganS23,Dibia23,abs-2309-10245}, such as Python, specifically tailored to data visualization. However, our approach differs from these existing methods by proposing to use LLMs as a comprehensive pipeline to generate queries for data visualization. This approach breaks the constraints of a singular declarative visualization language, offering a broader way for automatic data visualization.

%% file: conclusion.tex
\section{Conclusion}\label{section:cl}
Our work is a timely study of in-context learning for text-to-vis tasks in the era of large language models. We systematically study how to perform in-context learning for text-to-vis and propose a novel framework, \textsc{Prompt4Vis}, which includes an example mining module and a schema filtering module. Compared to the previous SOTA models and methods that randomly select examples for in-context learning, our model achieved significant results, demonstrating the immense potential of in-context learning in the field of text-to-vis even data science.

We also discover some interesting findings in this work, such as in the process of example selection for text-to-vis, aside from the similarity between examples, the influence and diversity of examples also play important roles. Moreover, for each test case, the full database schema may introduce redundant information. Hence, a good schema filtering module can reduce the irrelevant schema information and further enhance the performance. We believe that it is valuable to attempt and practice these approaches in other data science fields as well. 

%% file: sample-base.bbl

\begin{thebibliography}{45}


\ifx \showCODEN    \undefined \def \showCODEN     #1{\unskip}     \fi
\ifx \showDOI      \undefined \def \showDOI       #1{#1}\fi
\ifx \showISBNx    \undefined \def \showISBNx     #1{\unskip}     \fi
\ifx \showISBNxiii \undefined \def \showISBNxiii  #1{\unskip}     \fi
\ifx \showISSN     \undefined \def \showISSN      #1{\unskip}     \fi
\ifx \showLCCN     \undefined \def \showLCCN      #1{\unskip}     \fi
\ifx \shownote     \undefined \def \shownote      #1{#1}          \fi
\ifx \showarticletitle \undefined \def \showarticletitle #1{#1}   \fi
\ifx \showURL      \undefined \def \showURL       {\relax}        \fi
\providecommand\bibfield[2]{#2}
\providecommand\bibinfo[2]{#2}
\providecommand\natexlab[1]{#1}
\providecommand\showeprint[2][]{arXiv:#2}

\bibitem[Bang et~al\mbox{.}(2023)]%
        {abs-2302-04023}
\bibfield{author}{\bibinfo{person}{Yejin Bang}, \bibinfo{person}{Samuel Cahyawijaya}, \bibinfo{person}{Nayeon Lee}, \bibinfo{person}{Wenliang Dai}, \bibinfo{person}{Dan Su}, \bibinfo{person}{Bryan Wilie}, \bibinfo{person}{Holy Lovenia}, \bibinfo{person}{Ziwei Ji}, \bibinfo{person}{Tiezheng Yu}, \bibinfo{person}{Willy Chung}, \bibinfo{person}{Quyet~V. Do}, \bibinfo{person}{Yan Xu}, {and} \bibinfo{person}{Pascale Fung}.} \bibinfo{year}{2023}\natexlab{}.
\newblock \showarticletitle{A Multitask, Multilingual, Multimodal Evaluation of ChatGPT on Reasoning, Hallucination, and Interactivity}.
\newblock \bibinfo{journal}{\emph{CoRR}}  \bibinfo{volume}{abs/2302.04023} (\bibinfo{year}{2023}).
\newblock


\bibitem[Bar et~al\mbox{.}(2022)]%
        {BarGDGE22}
\bibfield{author}{\bibinfo{person}{Amir Bar}, \bibinfo{person}{Yossi Gandelsman}, \bibinfo{person}{Trevor Darrell}, \bibinfo{person}{Amir Globerson}, {and} \bibinfo{person}{Alexei~A. Efros}.} \bibinfo{year}{2022}\natexlab{}.
\newblock \showarticletitle{Visual Prompting via Image Inpainting}. In \bibinfo{booktitle}{\emph{Advances in Neural Information Processing Systems 35: Annual Conference on Neural Information Processing Systems 2022, NeurIPS 2022, New Orleans, LA, USA, November 28 - December 9, 2022}}.
\newblock


\bibitem[Brown et~al\mbox{.}(2020)]%
        {BrownMRSKDNSSAA20}
\bibfield{author}{\bibinfo{person}{Tom~B. Brown}, \bibinfo{person}{Benjamin Mann}, \bibinfo{person}{Nick Ryder}, \bibinfo{person}{Melanie Subbiah}, \bibinfo{person}{Jared Kaplan}, \bibinfo{person}{Prafulla Dhariwal}, \bibinfo{person}{Arvind Neelakantan}, \bibinfo{person}{Pranav Shyam}, \bibinfo{person}{Girish Sastry}, \bibinfo{person}{Amanda Askell}, \bibinfo{person}{Sandhini Agarwal}, \bibinfo{person}{Ariel Herbert{-}Voss}, \bibinfo{person}{Gretchen Krueger}, \bibinfo{person}{Tom Henighan}, \bibinfo{person}{Rewon Child}, \bibinfo{person}{Aditya Ramesh}, \bibinfo{person}{Daniel~M. Ziegler}, \bibinfo{person}{Jeffrey Wu}, \bibinfo{person}{Clemens Winter}, \bibinfo{person}{Christopher Hesse}, \bibinfo{person}{Mark Chen}, \bibinfo{person}{Eric Sigler}, \bibinfo{person}{Mateusz Litwin}, \bibinfo{person}{Scott Gray}, \bibinfo{person}{Benjamin Chess}, \bibinfo{person}{Jack Clark}, \bibinfo{person}{Christopher Berner}, \bibinfo{person}{Sam McCandlish}, \bibinfo{person}{Alec Radford}, \bibinfo{person}{Ilya Sutskever},
  {and} \bibinfo{person}{Dario Amodei}.} \bibinfo{year}{2020}\natexlab{}.
\newblock \showarticletitle{Language Models are Few-Shot Learners}. In \bibinfo{booktitle}{\emph{Advances in Neural Information Processing Systems 33: Annual Conference on Neural Information Processing Systems 2020, NeurIPS 2020, December 6-12, 2020, virtual}}.
\newblock


\bibitem[Chowdhery et~al\mbox{.}(2023)]%
        {ChowdheryNDBMRBCSGSSTMRBTSPRDHPBAI23}
\bibfield{author}{\bibinfo{person}{Aakanksha Chowdhery}, \bibinfo{person}{Sharan Narang}, \bibinfo{person}{Jacob Devlin}, \bibinfo{person}{Maarten Bosma}, \bibinfo{person}{Gaurav Mishra}, \bibinfo{person}{Adam Roberts}, \bibinfo{person}{Paul Barham}, \bibinfo{person}{Hyung~Won Chung}, \bibinfo{person}{Charles Sutton}, \bibinfo{person}{Sebastian Gehrmann}, \bibinfo{person}{Parker Schuh}, \bibinfo{person}{Kensen Shi}, \bibinfo{person}{Sasha Tsvyashchenko}, \bibinfo{person}{Joshua Maynez}, \bibinfo{person}{Abhishek Rao}, \bibinfo{person}{Parker Barnes}, \bibinfo{person}{Yi Tay}, \bibinfo{person}{Noam Shazeer}, \bibinfo{person}{Vinodkumar Prabhakaran}, \bibinfo{person}{Emily Reif}, \bibinfo{person}{Nan Du}, \bibinfo{person}{Ben Hutchinson}, \bibinfo{person}{Reiner Pope}, \bibinfo{person}{James Bradbury}, \bibinfo{person}{Jacob Austin}, \bibinfo{person}{Michael Isard}, \bibinfo{person}{Guy Gur{-}Ari}, \bibinfo{person}{Pengcheng Yin}, \bibinfo{person}{Toju Duke}, \bibinfo{person}{Anselm Levskaya},
  \bibinfo{person}{Sanjay Ghemawat}, \bibinfo{person}{Sunipa Dev}, \bibinfo{person}{Henryk Michalewski}, \bibinfo{person}{Xavier Garcia}, \bibinfo{person}{Vedant Misra}, \bibinfo{person}{Kevin Robinson}, \bibinfo{person}{Liam Fedus}, \bibinfo{person}{Denny Zhou}, \bibinfo{person}{Daphne Ippolito}, \bibinfo{person}{David Luan}, \bibinfo{person}{Hyeontaek Lim}, \bibinfo{person}{Barret Zoph}, \bibinfo{person}{Alexander Spiridonov}, \bibinfo{person}{Ryan Sepassi}, \bibinfo{person}{David Dohan}, \bibinfo{person}{Shivani Agrawal}, \bibinfo{person}{Mark Omernick}, \bibinfo{person}{Andrew~M. Dai}, \bibinfo{person}{Thanumalayan~Sankaranarayana Pillai}, \bibinfo{person}{Marie Pellat}, \bibinfo{person}{Aitor Lewkowycz}, \bibinfo{person}{Erica Moreira}, \bibinfo{person}{Rewon Child}, \bibinfo{person}{Oleksandr Polozov}, \bibinfo{person}{Katherine Lee}, \bibinfo{person}{Zongwei Zhou}, \bibinfo{person}{Xuezhi Wang}, \bibinfo{person}{Brennan Saeta}, \bibinfo{person}{Mark Diaz}, \bibinfo{person}{Orhan Firat},
  \bibinfo{person}{Michele Catasta}, \bibinfo{person}{Jason Wei}, \bibinfo{person}{Kathy Meier{-}Hellstern}, \bibinfo{person}{Douglas Eck}, \bibinfo{person}{Jeff Dean}, \bibinfo{person}{Slav Petrov}, {and} \bibinfo{person}{Noah Fiedel}.} \bibinfo{year}{2023}\natexlab{}.
\newblock \showarticletitle{PaLM: Scaling Language Modeling with Pathways}.
\newblock \bibinfo{journal}{\emph{J. Mach. Learn. Res.}}  \bibinfo{volume}{24} (\bibinfo{year}{2023}), \bibinfo{pages}{240:1--240:113}.
\newblock


\bibitem[Cui et~al\mbox{.}(2020)]%
        {CuiZWHCFZLZ20}
\bibfield{author}{\bibinfo{person}{Weiwei Cui}, \bibinfo{person}{Xiaoyu Zhang}, \bibinfo{person}{Yun Wang}, \bibinfo{person}{He Huang}, \bibinfo{person}{Bei Chen}, \bibinfo{person}{Lei Fang}, \bibinfo{person}{Haidong Zhang}, \bibinfo{person}{Jian{-}Guang Lou}, {and} \bibinfo{person}{Dongmei Zhang}.} \bibinfo{year}{2020}\natexlab{}.
\newblock \showarticletitle{Text-to-Viz: Automatic Generation of Infographics from Proportion-Related Natural Language Statements}.
\newblock \bibinfo{journal}{\emph{{IEEE} Trans. Vis. Comput. Graph.}} \bibinfo{volume}{26}, \bibinfo{number}{1} (\bibinfo{year}{2020}), \bibinfo{pages}{906--916}.
\newblock


\bibitem[Devlin et~al\mbox{.}(2019)]%
        {DevlinCLT19}
\bibfield{author}{\bibinfo{person}{Jacob Devlin}, \bibinfo{person}{Ming{-}Wei Chang}, \bibinfo{person}{Kenton Lee}, {and} \bibinfo{person}{Kristina Toutanova}.} \bibinfo{year}{2019}\natexlab{}.
\newblock \showarticletitle{{BERT:} Pre-training of Deep Bidirectional Transformers for Language Understanding}. In \bibinfo{booktitle}{\emph{Proceedings of the 2019 Conference of the North American Chapter of the Association for Computational Linguistics: Human Language Technologies, {NAACL-HLT} 2019, Minneapolis, MN, USA, June 2-7, 2019, Volume 1 (Long and Short Papers)}}, \bibfield{editor}{\bibinfo{person}{Jill Burstein}, \bibinfo{person}{Christy Doran}, {and} \bibinfo{person}{Thamar Solorio}} (Eds.). \bibinfo{publisher}{Association for Computational Linguistics}, \bibinfo{pages}{4171--4186}.
\newblock


\bibitem[Dibia(2023)]%
        {Dibia23}
\bibfield{author}{\bibinfo{person}{Victor Dibia}.} \bibinfo{year}{2023}\natexlab{}.
\newblock \showarticletitle{{LIDA:} {A} Tool for Automatic Generation of Grammar-Agnostic Visualizations and Infographics using Large Language Models}. In \bibinfo{booktitle}{\emph{Proceedings of the 61st Annual Meeting of the Association for Computational Linguistics: System Demonstrations, {ACL} 2023, Toronto, Canada, July 10-12, 2023}}, \bibfield{editor}{\bibinfo{person}{Danushka Bollegala}, \bibinfo{person}{Ruihong Huang}, {and} \bibinfo{person}{Alan Ritter}} (Eds.). \bibinfo{publisher}{Association for Computational Linguistics}, \bibinfo{pages}{113--126}.
\newblock


\bibitem[Dibia and Demiralp(2019)]%
        {DibiaD19}
\bibfield{author}{\bibinfo{person}{Victor Dibia} {and} \bibinfo{person}{{\c{C}}agatay Demiralp}.} \bibinfo{year}{2019}\natexlab{}.
\newblock \showarticletitle{Data2Vis: Automatic Generation of Data Visualizations Using Sequence-to-Sequence Recurrent Neural Networks}.
\newblock \bibinfo{journal}{\emph{{IEEE} Computer Graphics and Applications}} \bibinfo{volume}{39}, \bibinfo{number}{5} (\bibinfo{year}{2019}), \bibinfo{pages}{33--46}.
\newblock


\bibitem[Dong et~al\mbox{.}(2023)]%
        {abs-2307-07306}
\bibfield{author}{\bibinfo{person}{Xuemei Dong}, \bibinfo{person}{Chao Zhang}, \bibinfo{person}{Yuhang Ge}, \bibinfo{person}{Yuren Mao}, \bibinfo{person}{Yunjun Gao}, \bibinfo{person}{Lu Chen}, \bibinfo{person}{Jinshu Lin}, {and} \bibinfo{person}{Dongfang Lou}.} \bibinfo{year}{2023}\natexlab{}.
\newblock \showarticletitle{{C3:} Zero-shot Text-to-SQL with ChatGPT}.
\newblock \bibinfo{journal}{\emph{CoRR}}  \bibinfo{volume}{abs/2307.07306} (\bibinfo{year}{2023}).
\newblock


\bibitem[Gao et~al\mbox{.}(2023)]%
        {abs-2308-15363}
\bibfield{author}{\bibinfo{person}{Dawei Gao}, \bibinfo{person}{Haibin Wang}, \bibinfo{person}{Yaliang Li}, \bibinfo{person}{Xiuyu Sun}, \bibinfo{person}{Yichen Qian}, \bibinfo{person}{Bolin Ding}, {and} \bibinfo{person}{Jingren Zhou}.} \bibinfo{year}{2023}\natexlab{}.
\newblock \showarticletitle{Text-to-SQL Empowered by Large Language Models: {A} Benchmark Evaluation}.
\newblock \bibinfo{journal}{\emph{CoRR}}  \bibinfo{volume}{abs/2308.15363} (\bibinfo{year}{2023}).
\newblock
\urldef\tempurl%
\url{https://doi.org/10.48550/ARXIV.2308.15363}
\showDOI{\tempurl}
\showeprint[arXiv]{2308.15363}


\bibitem[Gao et~al\mbox{.}(2021)]%
        {GaoYC21}
\bibfield{author}{\bibinfo{person}{Tianyu Gao}, \bibinfo{person}{Xingcheng Yao}, {and} \bibinfo{person}{Danqi Chen}.} \bibinfo{year}{2021}\natexlab{}.
\newblock \showarticletitle{SimCSE: Simple Contrastive Learning of Sentence Embeddings}. In \bibinfo{booktitle}{\emph{Proceedings of the 2021 Conference on Empirical Methods in Natural Language Processing, {EMNLP} 2021, Virtual Event / Punta Cana, Dominican Republic, 7-11 November, 2021}}, \bibfield{editor}{\bibinfo{person}{Marie{-}Francine Moens}, \bibinfo{person}{Xuanjing Huang}, \bibinfo{person}{Lucia Specia}, {and} \bibinfo{person}{Scott~Wen{-}tau Yih}} (Eds.). \bibinfo{publisher}{Association for Computational Linguistics}, \bibinfo{pages}{6894--6910}.
\newblock


\bibitem[Hanrahan(2006)]%
        {Hanrahan06}
\bibfield{author}{\bibinfo{person}{Pat Hanrahan}.} \bibinfo{year}{2006}\natexlab{}.
\newblock \showarticletitle{VizQL: a language for query, analysis and visualization}. In \bibinfo{booktitle}{\emph{Proceedings of the {ACM} {SIGMOD} International Conference on Management of Data, Chicago, Illinois, USA, June 27-29, 2006}}, \bibfield{editor}{\bibinfo{person}{Surajit Chaudhuri}, \bibinfo{person}{Vagelis Hristidis}, {and} \bibinfo{person}{Neoklis Polyzotis}} (Eds.). \bibinfo{publisher}{{ACM}}, \bibinfo{pages}{721}.
\newblock


\bibitem[Ko et~al\mbox{.}(2023)]%
        {abs-2309-10245}
\bibfield{author}{\bibinfo{person}{Hyung{-}Kwon Ko}, \bibinfo{person}{Hyeon Jeon}, \bibinfo{person}{Gwanmo Park}, \bibinfo{person}{Dae~Hyun Kim}, \bibinfo{person}{Nam~Wook Kim}, \bibinfo{person}{Juho Kim}, {and} \bibinfo{person}{Jinwook Seo}.} \bibinfo{year}{2023}\natexlab{}.
\newblock \showarticletitle{Natural Language Dataset Generation Framework for Visualizations Powered by Large Language Models}.
\newblock \bibinfo{journal}{\emph{CoRR}}  \bibinfo{volume}{abs/2309.10245} (\bibinfo{year}{2023}).
\newblock
\urldef\tempurl%
\url{https://doi.org/10.48550/ARXIV.2309.10245}
\showDOI{\tempurl}
\showeprint[arXiv]{2309.10245}


\bibitem[Li et~al\mbox{.}(2018)]%
        {LiMSSZWZC18}
\bibfield{author}{\bibinfo{person}{Deqing Li}, \bibinfo{person}{Honghui Mei}, \bibinfo{person}{Yi Shen}, \bibinfo{person}{Shuang Su}, \bibinfo{person}{Wenli Zhang}, \bibinfo{person}{Junting Wang}, \bibinfo{person}{Ming Zu}, {and} \bibinfo{person}{Wei Chen}.} \bibinfo{year}{2018}\natexlab{}.
\newblock \showarticletitle{ECharts: {A} declarative framework for rapid construction of web-based visualization}.
\newblock \bibinfo{journal}{\emph{Vis. Informatics}} \bibinfo{volume}{2}, \bibinfo{number}{2} (\bibinfo{year}{2018}), \bibinfo{pages}{136--146}.
\newblock


\bibitem[Liu et~al\mbox{.}(2022)]%
        {LiuSZDCC22}
\bibfield{author}{\bibinfo{person}{Jiachang Liu}, \bibinfo{person}{Dinghan Shen}, \bibinfo{person}{Yizhe Zhang}, \bibinfo{person}{Bill Dolan}, \bibinfo{person}{Lawrence Carin}, {and} \bibinfo{person}{Weizhu Chen}.} \bibinfo{year}{2022}\natexlab{}.
\newblock \showarticletitle{What Makes Good In-Context Examples for GPT-3?}. In \bibinfo{booktitle}{\emph{Proceedings of Deep Learning Inside Out: The 3rd Workshop on Knowledge Extraction and Integration for Deep Learning Architectures, DeeLIO@ACL 2022, Dublin, Ireland and Online, May 27, 2022}}, \bibfield{editor}{\bibinfo{person}{Eneko Agirre}, \bibinfo{person}{Marianna Apidianaki}, {and} \bibinfo{person}{Ivan Vulic}} (Eds.). \bibinfo{publisher}{Association for Computational Linguistics}, \bibinfo{pages}{100--114}.
\newblock


\bibitem[Lu et~al\mbox{.}(2022)]%
        {LuBM0S22}
\bibfield{author}{\bibinfo{person}{Yao Lu}, \bibinfo{person}{Max Bartolo}, \bibinfo{person}{Alastair Moore}, \bibinfo{person}{Sebastian Riedel}, {and} \bibinfo{person}{Pontus Stenetorp}.} \bibinfo{year}{2022}\natexlab{}.
\newblock \showarticletitle{Fantastically Ordered Prompts and Where to Find Them: Overcoming Few-Shot Prompt Order Sensitivity}. In \bibinfo{booktitle}{\emph{Proceedings of the 60th Annual Meeting of the Association for Computational Linguistics (Volume 1: Long Papers), {ACL} 2022, Dublin, Ireland, May 22-27, 2022}}, \bibfield{editor}{\bibinfo{person}{Smaranda Muresan}, \bibinfo{person}{Preslav Nakov}, {and} \bibinfo{person}{Aline Villavicencio}} (Eds.). \bibinfo{publisher}{Association for Computational Linguistics}, \bibinfo{pages}{8086--8098}.
\newblock


\bibitem[Luo et~al\mbox{.}(2018a)]%
        {LuoQ0018}
\bibfield{author}{\bibinfo{person}{Yuyu Luo}, \bibinfo{person}{Xuedi Qin}, \bibinfo{person}{Nan Tang}, {and} \bibinfo{person}{Guoliang Li}.} \bibinfo{year}{2018}\natexlab{a}.
\newblock \showarticletitle{DeepEye: Towards Automatic Data Visualization}. In \bibinfo{booktitle}{\emph{34th {IEEE} International Conference on Data Engineering, {ICDE} 2018, Paris, France, April 16-19, 2018}}. \bibinfo{publisher}{{IEEE} Computer Society}, \bibinfo{pages}{101--112}.
\newblock


\bibitem[Luo et~al\mbox{.}(2018b)]%
        {LuoQ00W18}
\bibfield{author}{\bibinfo{person}{Yuyu Luo}, \bibinfo{person}{Xuedi Qin}, \bibinfo{person}{Nan Tang}, \bibinfo{person}{Guoliang Li}, {and} \bibinfo{person}{Xinran Wang}.} \bibinfo{year}{2018}\natexlab{b}.
\newblock \showarticletitle{DeepEye: Creating Good Data Visualizations by Keyword Search}. In \bibinfo{booktitle}{\emph{Proceedings of the 2018 International Conference on Management of Data, {SIGMOD} Conference 2018, Houston, TX, USA, June 10-15, 2018}}, \bibfield{editor}{\bibinfo{person}{Gautam Das}, \bibinfo{person}{Christopher~M. Jermaine}, {and} \bibinfo{person}{Philip~A. Bernstein}} (Eds.). \bibinfo{publisher}{{ACM}}, \bibinfo{pages}{1733--1736}.
\newblock


\bibitem[Luo et~al\mbox{.}(2021)]%
        {Luo00CLQ21}
\bibfield{author}{\bibinfo{person}{Yuyu Luo}, \bibinfo{person}{Nan Tang}, \bibinfo{person}{Guoliang Li}, \bibinfo{person}{Chengliang Chai}, \bibinfo{person}{Wenbo Li}, {and} \bibinfo{person}{Xuedi Qin}.} \bibinfo{year}{2021}\natexlab{}.
\newblock \showarticletitle{Synthesizing Natural Language to Visualization {(NL2VIS)} Benchmarks from {NL2SQL} Benchmarks}. In \bibinfo{booktitle}{\emph{{SIGMOD} '21: International Conference on Management of Data, Virtual Event, China, June 20-25, 2021}}, \bibfield{editor}{\bibinfo{person}{Guoliang Li}, \bibinfo{person}{Zhanhuai Li}, \bibinfo{person}{Stratos Idreos}, {and} \bibinfo{person}{Divesh Srivastava}} (Eds.). \bibinfo{publisher}{{ACM}}, \bibinfo{pages}{1235--1247}.
\newblock


\bibitem[Luo et~al\mbox{.}(2022)]%
        {LuoTLTCQ22}
\bibfield{author}{\bibinfo{person}{Yuyu Luo}, \bibinfo{person}{Nan Tang}, \bibinfo{person}{Guoliang Li}, \bibinfo{person}{Jiawei Tang}, \bibinfo{person}{Chengliang Chai}, {and} \bibinfo{person}{Xuedi Qin}.} \bibinfo{year}{2022}\natexlab{}.
\newblock \showarticletitle{Natural Language to Visualization by Neural Machine Translation}.
\newblock \bibinfo{journal}{\emph{{IEEE} Trans. Vis. Comput. Graph.}} \bibinfo{volume}{28}, \bibinfo{number}{1} (\bibinfo{year}{2022}), \bibinfo{pages}{217--226}.
\newblock


\bibitem[Lyu et~al\mbox{.}(2023)]%
        {abs-2301-13379}
\bibfield{author}{\bibinfo{person}{Qing Lyu}, \bibinfo{person}{Shreya Havaldar}, \bibinfo{person}{Adam Stein}, \bibinfo{person}{Li Zhang}, \bibinfo{person}{Delip Rao}, \bibinfo{person}{Eric Wong}, \bibinfo{person}{Marianna Apidianaki}, {and} \bibinfo{person}{Chris Callison{-}Burch}.} \bibinfo{year}{2023}\natexlab{}.
\newblock \showarticletitle{Faithful Chain-of-Thought Reasoning}.
\newblock \bibinfo{journal}{\emph{CoRR}}  \bibinfo{volume}{abs/2301.13379} (\bibinfo{year}{2023}).
\newblock
\urldef\tempurl%
\url{https://doi.org/10.48550/ARXIV.2301.13379}
\showDOI{\tempurl}


\bibitem[Maddigan and Susnjak(2023)]%
        {MaddiganS23}
\bibfield{author}{\bibinfo{person}{Paula Maddigan} {and} \bibinfo{person}{Teo Susnjak}.} \bibinfo{year}{2023}\natexlab{}.
\newblock \showarticletitle{Chat2VIS: Generating Data Visualizations via Natural Language Using ChatGPT, Codex and {GPT-3} Large Language Models}.
\newblock \bibinfo{journal}{\emph{{IEEE} Access}}  \bibinfo{volume}{11} (\bibinfo{year}{2023}), \bibinfo{pages}{45181--45193}.
\newblock


\bibitem[Moritz et~al\mbox{.}(2019)]%
        {MoritzWNLSHH19}
\bibfield{author}{\bibinfo{person}{Dominik Moritz}, \bibinfo{person}{Chenglong Wang}, \bibinfo{person}{Greg~L. Nelson}, \bibinfo{person}{Halden Lin}, \bibinfo{person}{Adam~M. Smith}, \bibinfo{person}{Bill Howe}, {and} \bibinfo{person}{Jeffrey Heer}.} \bibinfo{year}{2019}\natexlab{}.
\newblock \showarticletitle{Formalizing Visualization Design Knowledge as Constraints: Actionable and Extensible Models in Draco}.
\newblock \bibinfo{journal}{\emph{{IEEE} Trans. Vis. Comput. Graph.}} \bibinfo{volume}{25}, \bibinfo{number}{1} (\bibinfo{year}{2019}), \bibinfo{pages}{438--448}.
\newblock


\bibitem[Narayan et~al\mbox{.}(2022)]%
        {NarayanCOR22}
\bibfield{author}{\bibinfo{person}{Avanika Narayan}, \bibinfo{person}{Ines Chami}, \bibinfo{person}{Laurel~J. Orr}, {and} \bibinfo{person}{Christopher R{\'{e}}}.} \bibinfo{year}{2022}\natexlab{}.
\newblock \showarticletitle{Can Foundation Models Wrangle Your Data?}
\newblock \bibinfo{journal}{\emph{Proc. {VLDB} Endow.}} \bibinfo{volume}{16}, \bibinfo{number}{4} (\bibinfo{year}{2022}), \bibinfo{pages}{738--746}.
\newblock


\bibitem[Narechania et~al\mbox{.}(2021)]%
        {NarechaniaSS21}
\bibfield{author}{\bibinfo{person}{Arpit Narechania}, \bibinfo{person}{Arjun Srinivasan}, {and} \bibinfo{person}{John~T. Stasko}.} \bibinfo{year}{2021}\natexlab{}.
\newblock \showarticletitle{{NL4DV:} {A} Toolkit for Generating Analytic Specifications for Data Visualization from Natural Language Queries}.
\newblock \bibinfo{journal}{\emph{{IEEE} Trans. Vis. Comput. Graph.}} \bibinfo{volume}{27}, \bibinfo{number}{2} (\bibinfo{year}{2021}), \bibinfo{pages}{369--379}.
\newblock


\bibitem[Nguyen and Wong(2023)]%
        {abs-2302-11042}
\bibfield{author}{\bibinfo{person}{Tai Nguyen} {and} \bibinfo{person}{Eric Wong}.} \bibinfo{year}{2023}\natexlab{}.
\newblock \showarticletitle{In-context Example Selection with Influences}.
\newblock \bibinfo{journal}{\emph{CoRR}}  \bibinfo{volume}{abs/2302.11042} (\bibinfo{year}{2023}).
\newblock
\urldef\tempurl%
\url{https://doi.org/10.48550/ARXIV.2302.11042}
\showDOI{\tempurl}
\showeprint[arXiv]{2302.11042}


\bibitem[Nye et~al\mbox{.}(2021)]%
        {abs-2112-00114}
\bibfield{author}{\bibinfo{person}{Maxwell~I. Nye}, \bibinfo{person}{Anders~Johan Andreassen}, \bibinfo{person}{Guy Gur{-}Ari}, \bibinfo{person}{Henryk Michalewski}, \bibinfo{person}{Jacob Austin}, \bibinfo{person}{David Bieber}, \bibinfo{person}{David Dohan}, \bibinfo{person}{Aitor Lewkowycz}, \bibinfo{person}{Maarten Bosma}, \bibinfo{person}{David Luan}, \bibinfo{person}{Charles Sutton}, {and} \bibinfo{person}{Augustus Odena}.} \bibinfo{year}{2021}\natexlab{}.
\newblock \showarticletitle{Show Your Work: Scratchpads for Intermediate Computation with Language Models}.
\newblock \bibinfo{journal}{\emph{CoRR}}  \bibinfo{volume}{abs/2112.00114} (\bibinfo{year}{2021}).
\newblock
\showeprint[arXiv]{2112.00114}
\urldef\tempurl%
\url{https://arxiv.org/abs/2112.00114}
\showURL{%
\tempurl}


\bibitem[OpenAI(2022)]%
        {openai2022}
\bibfield{author}{\bibinfo{person}{OpenAI}.} \bibinfo{year}{2022}\natexlab{}.
\newblock \bibinfo{booktitle}{\emph{Introducing chatgpt}}.
\newblock
\urldef\tempurl%
\url{https://openai.com/blog/chatgpt.}
\showURL{%
\tempurl}


\bibitem[OpenAI(2023)]%
        {openai2023}
\bibfield{author}{\bibinfo{person}{OpenAI}.} \bibinfo{year}{2023}\natexlab{}.
\newblock \bibinfo{booktitle}{\emph{Gpt-4 technical report}}.
\newblock
\showeprint{2303.08774}~[cs.DL]


\bibitem[Pourreza and Rafiei(2023)]%
        {abs-2304-11015}
\bibfield{author}{\bibinfo{person}{Mohammadreza Pourreza} {and} \bibinfo{person}{Davood Rafiei}.} \bibinfo{year}{2023}\natexlab{}.
\newblock \showarticletitle{{DIN-SQL:} Decomposed In-Context Learning of Text-to-SQL with Self-Correction}.
\newblock \bibinfo{journal}{\emph{CoRR}}  \bibinfo{volume}{abs/2304.11015} (\bibinfo{year}{2023}).
\newblock
\urldef\tempurl%
\url{https://doi.org/10.48550/ARXIV.2304.11015}
\showDOI{\tempurl}
\showeprint[arXiv]{2304.11015}


\bibitem[Qian et~al\mbox{.}(2021)]%
        {QianRDKKMLC21}
\bibfield{author}{\bibinfo{person}{Xin Qian}, \bibinfo{person}{Ryan~A. Rossi}, \bibinfo{person}{Fan Du}, \bibinfo{person}{Sungchul Kim}, \bibinfo{person}{Eunyee Koh}, \bibinfo{person}{Sana Malik}, \bibinfo{person}{Tak~Yeon Lee}, {and} \bibinfo{person}{Joel Chan}.} \bibinfo{year}{2021}\natexlab{}.
\newblock \showarticletitle{Learning to Recommend Visualizations from Data}. In \bibinfo{booktitle}{\emph{{KDD} '21: The 27th {ACM} {SIGKDD} Conference on Knowledge Discovery and Data Mining, Virtual Event, Singapore, August 14-18, 2021}}, \bibfield{editor}{\bibinfo{person}{Feida Zhu}, \bibinfo{person}{Beng~Chin Ooi}, {and} \bibinfo{person}{Chunyan Miao}} (Eds.). \bibinfo{publisher}{{ACM}}, \bibinfo{pages}{1359--1369}.
\newblock


\bibitem[Qin et~al\mbox{.}(2023)]%
        {QinZ0CYY23}
\bibfield{author}{\bibinfo{person}{Chengwei Qin}, \bibinfo{person}{Aston Zhang}, \bibinfo{person}{Zhuosheng Zhang}, \bibinfo{person}{Jiaao Chen}, \bibinfo{person}{Michihiro Yasunaga}, {and} \bibinfo{person}{Diyi Yang}.} \bibinfo{year}{2023}\natexlab{}.
\newblock \showarticletitle{Is ChatGPT a General-Purpose Natural Language Processing Task Solver?}. In \bibinfo{booktitle}{\emph{Proceedings of the 2023 Conference on Empirical Methods in Natural Language Processing, {EMNLP} 2023, Singapore, December 6-10, 2023}}, \bibfield{editor}{\bibinfo{person}{Houda Bouamor}, \bibinfo{person}{Juan Pino}, {and} \bibinfo{person}{Kalika Bali}} (Eds.). \bibinfo{publisher}{Association for Computational Linguistics}, \bibinfo{pages}{1339--1384}.
\newblock


\bibitem[Rubin et~al\mbox{.}(2022)]%
        {RubinHB22}
\bibfield{author}{\bibinfo{person}{Ohad Rubin}, \bibinfo{person}{Jonathan Herzig}, {and} \bibinfo{person}{Jonathan Berant}.} \bibinfo{year}{2022}\natexlab{}.
\newblock \showarticletitle{Learning To Retrieve Prompts for In-Context Learning}. In \bibinfo{booktitle}{\emph{Proceedings of the 2022 Conference of the North American Chapter of the Association for Computational Linguistics: Human Language Technologies, {NAACL} 2022, Seattle, WA, United States, July 10-15, 2022}}, \bibfield{editor}{\bibinfo{person}{Marine Carpuat}, \bibinfo{person}{Marie{-}Catherine de~Marneffe}, {and} \bibinfo{person}{Iv{\'{a}}n Vladimir~Meza Ru{\'{\i}}z}} (Eds.). \bibinfo{publisher}{Association for Computational Linguistics}, \bibinfo{pages}{2655--2671}.
\newblock


\bibitem[Satyanarayan et~al\mbox{.}(2017)]%
        {Vega}
\bibfield{author}{\bibinfo{person}{Arvind Satyanarayan}, \bibinfo{person}{Dominik Moritz}, \bibinfo{person}{Kanit Wongsuphasawat}, {and} \bibinfo{person}{Jeffrey Heer}.} \bibinfo{year}{2017}\natexlab{}.
\newblock \showarticletitle{Vega-Lite: {A} Grammar of Interactive Graphics}.
\newblock \bibinfo{journal}{\emph{{IEEE} Trans. Vis. Comput. Graph.}} \bibinfo{volume}{23}, \bibinfo{number}{1} (\bibinfo{year}{2017}), \bibinfo{pages}{341--350}.
\newblock


\bibitem[Savvides et~al\mbox{.}(2019)]%
        {SavvidesHOP19}
\bibfield{author}{\bibinfo{person}{Rafael Savvides}, \bibinfo{person}{Andreas Henelius}, \bibinfo{person}{Emilia Oikarinen}, {and} \bibinfo{person}{Kai Puolam{\"{a}}ki}.} \bibinfo{year}{2019}\natexlab{}.
\newblock \showarticletitle{Significance of Patterns in Data Visualisations}. In \bibinfo{booktitle}{\emph{Proceedings of the 25th {ACM} {SIGKDD} International Conference on Knowledge Discovery {\&} Data Mining, {KDD} 2019, Anchorage, AK, USA, August 4-8, 2019}}, \bibfield{editor}{\bibinfo{person}{Ankur Teredesai}, \bibinfo{person}{Vipin Kumar}, \bibinfo{person}{Ying Li}, \bibinfo{person}{R{\'{o}}mer Rosales}, \bibinfo{person}{Evimaria Terzi}, {and} \bibinfo{person}{George Karypis}} (Eds.). \bibinfo{publisher}{{ACM}}, \bibinfo{pages}{1509--1517}.
\newblock


\bibitem[Siddiqui et~al\mbox{.}(2016)]%
        {SiddiquiKLKP16}
\bibfield{author}{\bibinfo{person}{Tarique Siddiqui}, \bibinfo{person}{Albert Kim}, \bibinfo{person}{John Lee}, \bibinfo{person}{Karrie Karahalios}, {and} \bibinfo{person}{Aditya~G. Parameswaran}.} \bibinfo{year}{2016}\natexlab{}.
\newblock \showarticletitle{Effortless Data Exploration with zenvisage: An Expressive and Interactive Visual Analytics System}.
\newblock \bibinfo{journal}{\emph{Proc. {VLDB} Endow.}} \bibinfo{volume}{10}, \bibinfo{number}{4} (\bibinfo{year}{2016}), \bibinfo{pages}{457--468}.
\newblock


\bibitem[Song et~al\mbox{.}(2022)]%
        {SongZWJ22}
\bibfield{author}{\bibinfo{person}{Yuanfeng Song}, \bibinfo{person}{Xuefang Zhao}, \bibinfo{person}{Raymond~Chi{-}Wing Wong}, {and} \bibinfo{person}{Di Jiang}.} \bibinfo{year}{2022}\natexlab{}.
\newblock \showarticletitle{RGVisNet: {A} Hybrid Retrieval-Generation Neural Framework Towards Automatic Data Visualization Generation}. In \bibinfo{booktitle}{\emph{{KDD} '22: The 28th {ACM} {SIGKDD} Conference on Knowledge Discovery and Data Mining, Washington, DC, USA, August 14 - 18, 2022}}, \bibfield{editor}{\bibinfo{person}{Aidong Zhang} {and} \bibinfo{person}{Huzefa Rangwala}} (Eds.). \bibinfo{publisher}{{ACM}}, \bibinfo{pages}{1646--1655}.
\newblock


\bibitem[Vartak et~al\mbox{.}(2016)]%
        {VartakHSMP16}
\bibfield{author}{\bibinfo{person}{Manasi Vartak}, \bibinfo{person}{Silu Huang}, \bibinfo{person}{Tarique Siddiqui}, \bibinfo{person}{Samuel Madden}, {and} \bibinfo{person}{Aditya~G. Parameswaran}.} \bibinfo{year}{2016}\natexlab{}.
\newblock \showarticletitle{Towards Visualization Recommendation Systems}.
\newblock \bibinfo{journal}{\emph{{SIGMOD} Rec.}} \bibinfo{volume}{45}, \bibinfo{number}{4} (\bibinfo{year}{2016}), \bibinfo{pages}{34--39}.
\newblock


\bibitem[Vaswani et~al\mbox{.}(2017)]%
        {VaswaniSPUJGKP17}
\bibfield{author}{\bibinfo{person}{Ashish Vaswani}, \bibinfo{person}{Noam Shazeer}, \bibinfo{person}{Niki Parmar}, \bibinfo{person}{Jakob Uszkoreit}, \bibinfo{person}{Llion Jones}, \bibinfo{person}{Aidan~N. Gomez}, \bibinfo{person}{Lukasz Kaiser}, {and} \bibinfo{person}{Illia Polosukhin}.} \bibinfo{year}{2017}\natexlab{}.
\newblock \showarticletitle{Attention is All you Need}. In \bibinfo{booktitle}{\emph{Advances in Neural Information Processing Systems 30: Annual Conference on Neural Information Processing Systems 2017, December 4-9, 2017, Long Beach, CA, {USA}}}. \bibinfo{pages}{5998--6008}.
\newblock


\bibitem[Wang et~al\mbox{.}(2020)]%
        {WangSLPR20}
\bibfield{author}{\bibinfo{person}{Bailin Wang}, \bibinfo{person}{Richard Shin}, \bibinfo{person}{Xiaodong Liu}, \bibinfo{person}{Oleksandr Polozov}, {and} \bibinfo{person}{Matthew Richardson}.} \bibinfo{year}{2020}\natexlab{}.
\newblock \showarticletitle{{RAT-SQL:} Relation-Aware Schema Encoding and Linking for Text-to-SQL Parsers}. In \bibinfo{booktitle}{\emph{Proceedings of the 58th Annual Meeting of the Association for Computational Linguistics, {ACL} 2020, Online, July 5-10, 2020}}. \bibinfo{publisher}{Association for Computational Linguistics}, \bibinfo{pages}{7567--7578}.
\newblock


\bibitem[Wang et~al\mbox{.}(2024)]%
        {WangTL24}
\bibfield{author}{\bibinfo{person}{Chenglong Wang}, \bibinfo{person}{John Thompson}, {and} \bibinfo{person}{Bongshin Lee}.} \bibinfo{year}{2024}\natexlab{}.
\newblock \showarticletitle{Data Formulator: AI-Powered Concept-Driven Visualization Authoring}.
\newblock \bibinfo{journal}{\emph{{IEEE} Trans. Vis. Comput. Graph.}} \bibinfo{volume}{30}, \bibinfo{number}{1} (\bibinfo{year}{2024}), \bibinfo{pages}{1128--1138}.
\newblock


\bibitem[Wang et~al\mbox{.}(2023)]%
        {WangZWLW23}
\bibfield{author}{\bibinfo{person}{Lei Wang}, \bibinfo{person}{Songheng Zhang}, \bibinfo{person}{Yun Wang}, \bibinfo{person}{Ee{-}Peng Lim}, {and} \bibinfo{person}{Yong Wang}.} \bibinfo{year}{2023}\natexlab{}.
\newblock \showarticletitle{LLM4Vis: Explainable Visualization Recommendation using ChatGPT}. In \bibinfo{booktitle}{\emph{Proceedings of the 2023 Conference on Empirical Methods in Natural Language Processing: {EMNLP} 2023 - Industry Track, Singapore, December 6-10, 2023}}, \bibfield{editor}{\bibinfo{person}{Mingxuan Wang} {and} \bibinfo{person}{Imed Zitouni}} (Eds.). \bibinfo{publisher}{Association for Computational Linguistics}, \bibinfo{pages}{675--692}.
\newblock


\bibitem[Wei et~al\mbox{.}(2022)]%
        {Wei0SBIXCLZ22}
\bibfield{author}{\bibinfo{person}{Jason Wei}, \bibinfo{person}{Xuezhi Wang}, \bibinfo{person}{Dale Schuurmans}, \bibinfo{person}{Maarten Bosma}, \bibinfo{person}{Brian Ichter}, \bibinfo{person}{Fei Xia}, \bibinfo{person}{Ed~H. Chi}, \bibinfo{person}{Quoc~V. Le}, {and} \bibinfo{person}{Denny Zhou}.} \bibinfo{year}{2022}\natexlab{}.
\newblock \showarticletitle{Chain-of-Thought Prompting Elicits Reasoning in Large Language Models}. In \bibinfo{booktitle}{\emph{Advances in Neural Information Processing Systems 35: Annual Conference on Neural Information Processing Systems 2022, NeurIPS 2022, New Orleans, LA, USA, November 28 - December 9, 2022}}, \bibfield{editor}{\bibinfo{person}{Sanmi Koyejo}, \bibinfo{person}{S.~Mohamed}, \bibinfo{person}{A.~Agarwal}, \bibinfo{person}{Danielle Belgrave}, \bibinfo{person}{K.~Cho}, {and} \bibinfo{person}{A.~Oh}} (Eds.).
\newblock


\bibitem[Wickham(2009)]%
        {ggplot2}
\bibfield{author}{\bibinfo{person}{Hadley Wickham}.} \bibinfo{year}{2009}\natexlab{}.
\newblock \bibinfo{booktitle}{\emph{ggplot2 - Elegant Graphics for Data Analysis}}.
\newblock \bibinfo{publisher}{Springer}.
\newblock
\showISBNx{978-0-387-98140-6}
\urldef\tempurl%
\url{https://doi.org/10.1007/978-0-387-98141-3}
\showDOI{\tempurl}


\bibitem[Zhang et~al\mbox{.}(2023)]%
        {zhang2023VisualPromptRetrieval}
\bibfield{author}{\bibinfo{person}{Yuanhan Zhang}, \bibinfo{person}{Kaiyang Zhou}, {and} \bibinfo{person}{Ziwei Liu}.} \bibinfo{year}{2023}\natexlab{}.
\newblock \showarticletitle{What Makes Good Examples for Visual In-Context Learning?}
\newblock


\end{thebibliography}
